\documentclass[aip,reprint]{revtex4-1}

\makeatletter
\def\@email#1#2{%
 \endgroup
 \patchcmd{\titleblock@produce}
  {\frontmatter@RRAPformat}
  {\frontmatter@RRAPformat{\produce@RRAP{*#1\href{mailto:#2}{#2}}}\frontmatter@RRAPformat}
  {}{}
}%
\makeatother

\usepackage[utf8]{inputenc}
\usepackage[T1]{fontenc}
\usepackage{amsmath}
\usepackage{siunitx}
\usepackage[labelsep=period]{caption}
\usepackage[table,xcdraw]{xcolor}
\usepackage{framed}
\usepackage[normalem]{ulem}
\usepackage{amsfonts}
\usepackage{enumerate}
\usepackage{listings}
\usepackage{float}
\usepackage{textcomp}
\usepackage{makeidx}
\usepackage{mathtools}
\usepackage{subfigure}
\usepackage{verbatim}
\usepackage{caption}
\usepackage{hyperref}
\usepackage{mathptmx}
\usepackage{etoolbox}
\usepackage{amssymb}
\usepackage{amsthm}
\usepackage{graphicx}
\usepackage{bm}
\begin{document}

\title{Investigation of rotational augmentation mechanisms on wind turbine blade sections based on Quasi-3D simulations}

\author{Pedro Rodrigues}
    \altaffiliation{Aeronautical Engineer.}
\author{Elmer Gennaro}
    \altaffiliation{Assistant Professor, Department of Aeronautical Engineering.}
\author{Daniel Souza\textsuperscript{\text{*}}}
    \email{daniel.s.souza@unesp.br}
    \altaffiliation{Assistant Professor, Department of Aeronautical Engineering.}
\affiliation{\textsuperscript{1}Department of Aeronautical Engineering, São Paulo State University (UNESP), Campus of São João da Boa Vista, Av. Profª Isette Corrêa Fontão, 505 - Jardim das Flores, São João da Boa Vista, 13876-750, SP, Brazil}

\date{\today}

\begin{abstract}
Rotation has been recognized for its ability to decrease the extent of separate flow regions on the inner sections of horizontal-axis wind turbine blades. This alteration is often linked to centrifugal pumping. Nevertheless, studies focused on insect flight conditions revealed a significant contribution of the Coriolis acceleration to the balance of vorticity within the leading-edge vortex (LEV). Despite this insight, it remains unexplored whether this phenomenon contributes to the observed rotational augmentation in wind turbines.
Our study employed quasi-3D simulations to delve into the impact of Coriolis acceleration on the balance of span-wise vorticity within the region of separate flow along the upper surface of wind turbine blade sections. Our results allowed the identification of two different mechanisms of rotational augmentation, one associated to the height reduction of the region of separate flow and the other associated to the stabilization of the LEV. The former occurred when a trailing-edge separation was observed and the later when the flow featured a leading-edge separation. Within the range of conditions considered, the Coriolis acceleration neutralized up to about 40\% of the span-wise vorticity flowing in from the boundary layer to feed the LEV.
\end{abstract}

\maketitle

\section{Introduction} \label{sec:introduction}

Sections of rotating wings close to the rub exhibit an enhanced lift-generation capability when compared to their non-rotating counterparts under similar conditions of local angle of attack and Reynolds number \citep{dwyer:1970,ronsten:1992}. One of the mechanisms contributing to this rotational augmentation is the increased resistance of the boundary layer to separation, due to the chord-wise component of the Coriolis acceleration \citep{du:2000}.
However, large lift increase is commonly observed particularly when boundary-layer separation occurs \citep{mauro2:2017,sicot:2008,schreck:2007}.

Rotating blades are subject to a radial gradient of dynamic pressure of the incoming flow which induce a static-pressure gradient within the boundary layer, and thus may lead to a secondary flow towards the blade tip \citep{jardin:2014}.
Furthermore, the centrifugal acceleration in a reference frame moving with the blade is proportional to the distance from the axis of rotation \citep{corten:2001,lindenburg:2003}, further amplifying the flow towards the tip.
This so-called centrifugal pumping is capable of transporting vorticity away from inboard sections.

By comparing the flow patterns over an inboard section of an horizontal-axis wind turbine (HAWT) blade with the two-dimensional flow field for the same angle of attack (AoA), \citet{bangga:2017b} demonstrated a significant reduction in the recirculating region volume for the blade section. They attributed this reduction to centrifugal pumping and argued that it represents a weakening of the decambering effect, which contributes to lift augmentation. 

Studies considering the influence of rotation on the three-dimensional boundary-layer equations, as explored by \citet{dumitrescu:2007}, have suggested the emergence of a span-wise standing vortex on the suction surface of the blade. Their findings indicate that the vortex arises from sections where the chord-to-radius ratio falls within the range of 0.5 to 1.0, contingent upon the imposed pressure gradient. Analyzing unsteady pressure measurements from the inboard blade sections of a HAWT, \citet{schreck:2007} noted a substantial increase of the lift coefficient, compared to two-dimensional flow, correlated with the presence of a stationary vortex over the blade surface, revealed by the presence of two loci of intense pressure fluctuations. \citet{mauro2:2017} reached similar conclusions through three-dimensional simulations of a small-scale HAWT.

\citet{jardin:2014} associates the radial pressure and centrifugal force gradients with the stabilization of the leading-edge vortex (LEV) in insect flights.
The transport of vorticity towards the blade tip caused by these effects limits the growth of the LEV, delaying its breakdown to higher angles of attack, similar to what is observed in delta wings \citep{polhamus:1966}.
Nonetheless, there are evidence that other mechanisms may be relevant for the stabilization of the LEV on insect wings \citep{wojcik:2014,jardin:2017,werner:2019}.
Similarly, results by \citet{souza:2020} suggests that other physical mechanisms besides the centrifugal pumping may contribute to the height reduction of recirculating-flow region on HAWT blade sections.

An alternative mechanism was proposed by \citet{werner:2019}.
They showed that the radial tilting of the planetary vorticity, i.e. the contribution of the azimuthal Coriolis acceleration to the span-wise vorticity transport equation, opposes the vorticity of the LEV.
Considering representative conditions of fly wing operations, they showed that this mechanism tends to restrict the growth of the LEV, and thus, its breakdown.

The work described here aims at investigating the flow field around an HAWT blade section under different conditions and assessing the significance of the term related to radial planetary vorticity tilting to the balance of the radial vorticity within the region of separate flow on the upper surface of an HAWT section using numerical simulations. The study employed a quasi-3D model, inspired by the work of \citet{chaviaropoulos:2000}, and examines in detail three distinct angles of attack. Each angle of attack is evaluated in both two-dimensional (non-rotating) flow and under conditions corresponding to a section of an HAWT.

\section{Methodology}

\subsection{\textit{Quasi-3D simulations}}
\label{subsection:quasi-3D-simulations}

\begin{figure}
\begin{center}
\subfigure[\label{subfig:turbine}]{\includegraphics[width=0.50\columnwidth]{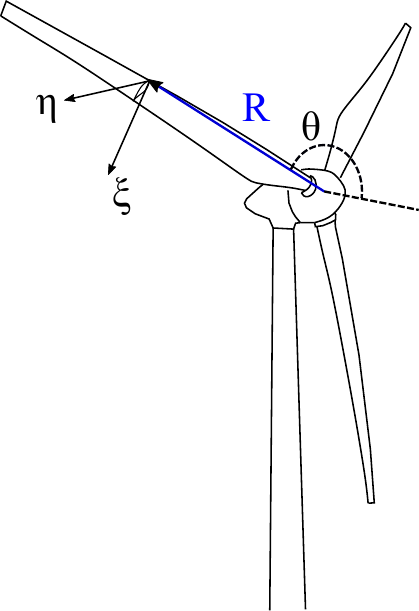}}
\subfigure[\label{subfig:section}]{\includegraphics[width=0.85\columnwidth]{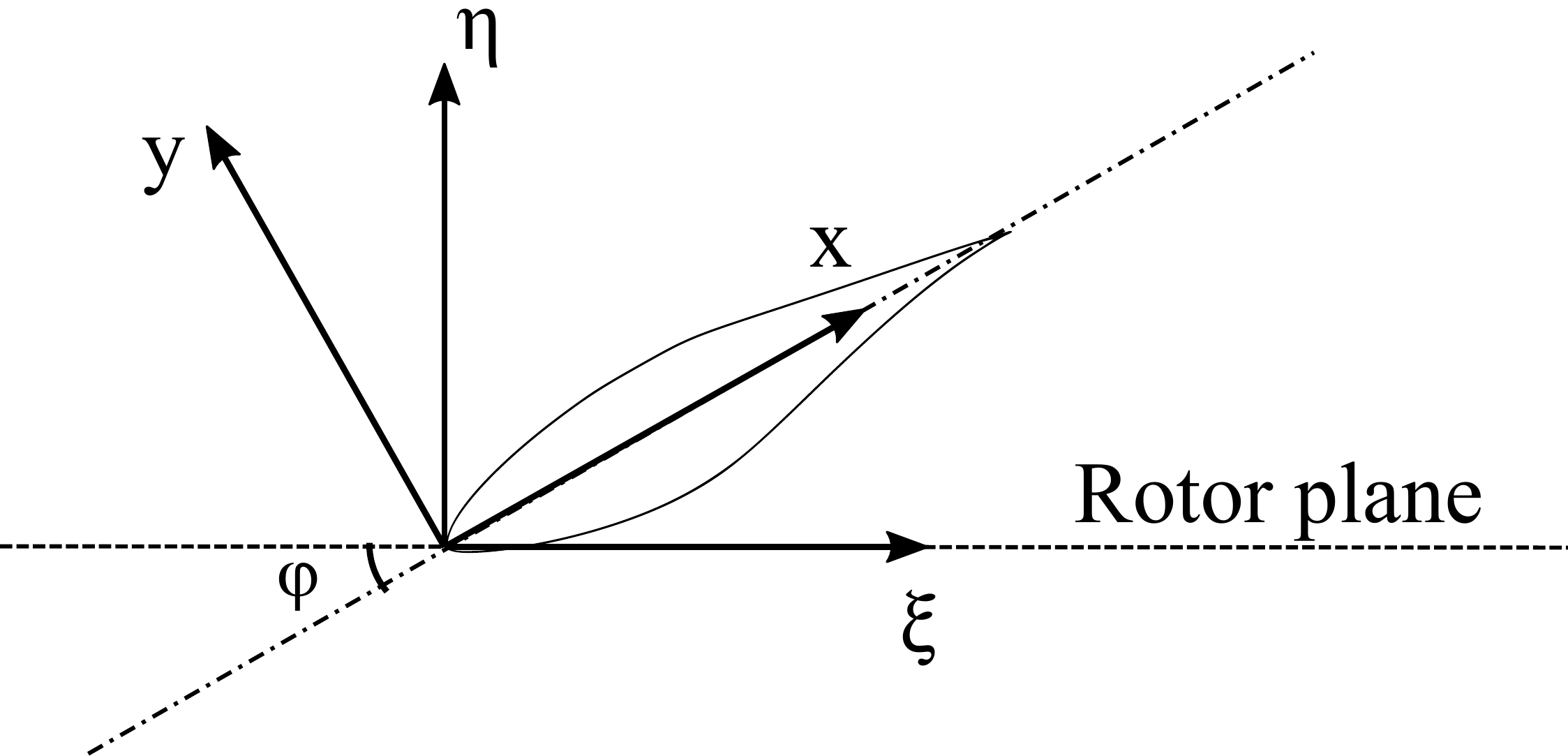}}
\caption{Schematic representation of (a) an horizontal-axis wind turbine (HAWT) highlighting a particular blade section and (b) detail of the section of radial position $R$ and relevant coordinate systems. \label{fig:schematic_turbine}}
\end{center}
\end{figure}

Consider the incompressible flow over a section at radial position $R$ of a blade schematically represented in Figure \ref{fig:schematic_turbine}.
The flow governing equations on a reference frame moving with the section in the curvilinear coordinates $(\xi, \eta, r)$ can be written as

\begin{equation}
\frac{\partial u_r}{\partial r} + \frac{u_r}{R} + \frac{\partial u_{\xi}}{\partial \xi} + \frac{\partial u_{\eta}}{\partial \eta}=0, \nonumber
\end{equation}

\begin{eqnarray*}
\frac{\partial u_r}{\partial t} &+& \frac{\partial (u_r^2)}{\partial r} + \frac{\partial (u_r u_{\xi})}{\partial \xi} + \frac{\partial (u_r u_{\eta})}{\partial \eta} + \frac{\partial \tilde{p}}{\partial r} - \frac{\partial \tau_{rr}}{\partial r} + \\ &-& \frac{\tau_{rr}}{R} - \frac{\partial \tau_{\xi r}}{\partial \xi} - \frac{\partial \tau_{\eta r}}{\partial \eta} + \frac{\tau_{\xi \xi}}{R} = \frac{u_{\xi}^2}{R} - \frac{u_{r}^2}{R} + \\ &-& 2 \Omega u_{\xi} + \Omega^2 R,
\end{eqnarray*}

\begin{eqnarray*}
\frac{\partial u_{\xi}}{\partial t} &+& \frac{\partial (u_r u_{\xi})}{\partial r} + \frac{\partial (u_{\xi}^2)}{\partial \xi} + \frac{\partial (u_{\xi} u_{\eta})}{\partial \eta} + \frac{\partial \tilde{p}}{\partial \xi} - \frac{\partial \tau_{r \xi}}{\partial r} + \\ &-& \frac{\partial \tau_{\xi \xi}}{\partial \xi} - \frac{\partial \tau_{\eta \xi}}{\partial \eta} - 2 \frac{\tau_{\xi r}}{R} = - 2 \frac{u_{\xi} u_r}{R} + 2 \Omega u_{r},
\end{eqnarray*}

\begin{eqnarray*}
\frac{\partial u_{\eta}}{\partial t} + \frac{\partial (u_r u_{\eta})}{\partial r} + \frac{\partial (u_{\xi} u_{\eta})}{\partial \xi} + \frac{\partial (u_{\eta}^2)}{\partial \eta} + \frac{\partial \tilde{p}}{\partial \eta} - \frac{\partial \tau_{r \eta}}{\partial r} + && \\ - \frac{\partial \tau_{\xi \eta}}{\partial \xi} - \frac{\partial \tau_{\eta \eta}}{\partial \eta} - \frac{\tau_{r \eta}}{R} = - \frac{u_{\eta} u_r}{R},
\end{eqnarray*}

\noindent where $t$ denotes time, $\Omega$ is the absolute rotor angular speed, $\tilde{p}=p/\rho$, $p$ is the static pressure and $\rho$ the fluid density.
Note that the coordinate $\xi$ is equal to $R \theta$, being $\theta$ the azimuthal coordinate along the rotor plane.
The tensions are described as follows

\begin{equation}
\tau_{r r} = \nu \left[ 2 \frac{\partial u_r}{\partial r}  \right] \nonumber
\end{equation}

\begin{equation}
\tau_{\xi \xi} = \nu \left[ 2 \left(\frac{\partial u_{\xi}}{\partial \xi} + \frac{u_r}{R} \right)  \right] \nonumber
\end{equation}

\begin{equation}
\tau_{\eta \eta} = \nu \left[ 2 \frac{\partial u_{\eta}}{\partial \eta}  \right] \nonumber
\end{equation}

\begin{equation}
\tau_{\xi r} = \nu \left[ \frac{\partial u_{r}}{\partial \xi} - \frac{u_{\xi}}{R} + \frac{\partial u_{\xi}}{\partial r} \right] \nonumber
\end{equation}

\begin{equation}
\tau_{\xi \eta} = \nu \left[ \frac{\partial u_{\eta}}{\partial \xi} + \frac{\partial u_{\xi}}{\partial \eta} \right] \nonumber
\end{equation}

\begin{equation}
\tau_{\eta r} = \nu \left[ \frac{\partial u_{\eta}}{\partial r} + \frac{\partial u_r}{\partial \eta} \right] \nonumber
\end{equation}

\noindent where $\nu$ is the kinematic viscosity.

The curvature terms were neglected, except the ones proportional to $u_{\xi}/R$, since $u_{\xi}$ is proportional to $R$.
Moreover, we neglected the radial derivatives.
Although this is a rather radical approximation, the resulting model is similar to the one developed by \citet{chaviaropoulos:2000} to generate the data employed to calibrate their broadly used stall delay model.
In the coordinates aligned with the section chord $(x,y)$ the simplified system of governing equations reads

\begin{equation}
\nabla \cdot \mathbf{u} = 0,
\label{eq:continuity_xy}
\end{equation}

\begin{equation}
\frac{\partial \mathbf{u}}{\partial t} + \nabla \cdot (\mathbf{u} \mathbf{u}) = \nabla \tilde{p} + \nabla \cdot \mathbf{\tilde{\tau}} + \mathbf{S},
\label{eq:xyMomentum}
\end{equation}

\begin{equation}
\frac{\partial w}{\partial t} + \nabla \cdot (w \mathbf{u}) = \nabla \cdot (\nu \nabla w) + \frac{u_{\xi}^2}{R} - 2 \Omega u_{\xi} + \Omega^2 R,
\label{eq:zMomentum}
\end{equation}

\noindent where $\mathbf{u}=(u,v)$ is the velocity vector in the $xy$ plane, $w$ is the radial velocity component, $[\mathbf{\tilde{\tau}}]$ is the deviatoric stress tensor in two dimensions, which, with the simplifications adopted, resembles the tensor in rectangular coordinates.
Likewise, the divergent and gradient operators correspond to the two-dimensional ones in rectangular coordinates $xy$.
The extra term in equation \ref{eq:xyMomentum} is described by

\begin{equation}
\mathbf{S} = \left( \frac{u_{\xi} w \cos\varphi}{R} + 2 \Omega w \cos\varphi, -\frac{u_{\xi} w \sin\varphi}{R} - 2 \Omega w \sin\varphi \right),
\end{equation}

\noindent and $u_{\xi}=(u,v) \cdot (\cos\varphi,-\sin\varphi)$, where $\varphi$ is the local pitch angle (see figure \ref{subfig:section}).

Equations \ref{eq:continuity_xy} through \ref{eq:zMomentum} were solved using the two-dimensional, incompressible, fully turbulent formulation implemented in the commercial package Ansys FLUENT \textsuperscript{\textregistered}.
The equation for the radial momentum was inserted as an additional scalar transport equation and the term $\mathbf{S}$ in the right-hand side of equation \ref{eq:xyMomentum} was inserted as a source term in the momentum equations.

Second-order discretization schemes were used: linear-upwind scheme for the convective terms and linear central-differences scheme for the viscous terms.
The solution algorithm SIMPLE \citep{patankar:1980} was employed to couple the pressure and velocity fields.
Due to its success modeling flow with separation \citep{mauro1:2017,wang:2010}, the SST $k-\omega$ \citep{menter:1994} turbulence model was adopted.
On the outer boundary of the domain, the prescribed velocity ensured attainment of the desired Reynolds number and angle of attack.
Notably, the radial velocity component was held at zero at the far field.
Stationary wall with no-slip condition on the airfoil surface was adopted.

The flow over the blade section governed by equations \ref{eq:continuity_xy} through \ref{eq:zMomentum} can be characterized by the Rossby number $Ro=U_{\infty}/(c \Omega)$ and the chord-to-radius ratio $c/r$, besides the chord-based Reynolds number $Re=\rho U_{\infty} c/\mu$ and the angle of attack $\alpha$.
The choice of the values of $\alpha$, $c/r$ and $Ro$ of the simulations was based on a study about the operating range of the wind turbine blade sections used in sequence H of the experiment \textit{Phase VI} from the NREL \cite{hand:2001} obtained for wind speed equal to 15m/s, whose variation along the sections comprises angles of attack between 0 and 48 degrees, ratios between local chord and radial position between 0 and 0.55, and Rossby numbers between 2 and 16. Thus, the values taken as a basis included the angles of attack of 10, 15, 27 and 45 degrees, $c/r$ ratios of 0.0625, 0.3, 0.45 and 0.55, and Rossby numbers of 2, 3, 3.5, 7, 9 and 16, thus running a total of 20 quasi-3D CFD simulations.

For angles of attack above the stall, equal to 27 and 45 degrees, unsteady simulations were carried out. Since the unsteady simulation time tends to be longer than that observed in steady state, an important step consisted in carrying out a study of the computational cost, measured in hours, for different time steps. A maximum number of iterations per time step equal to 40 was then considered, and the solution was iterated until reaching a statistically converged temporal response.

\subsection{\textit{Computational model}}

The study considered the flow over the 21\% thick S809 airfoil, designed by \citet{somers:1997}.
This geometry was designed to have low drag between the condition corresponding to a lift coefficient $C_l$=0.2 and $C_l$=0.8, have a minimum pitching-moment $C_m$ of -0.05 and a stall angle of attack insensible to the presence of roughness near the leading edge.

\begin{figure*}
                \centering
                \subfigure[\label{subfig:O-grid}]{
                \includegraphics[width=0.31\textwidth]{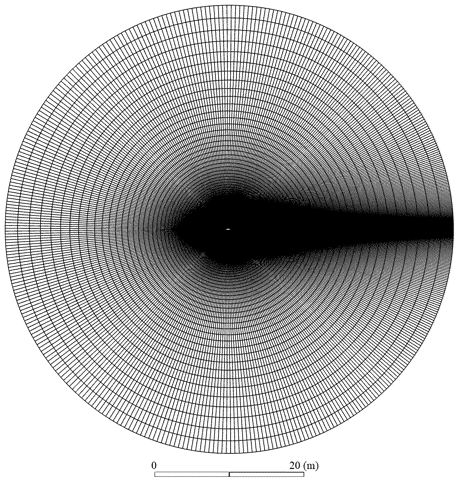}}
                \subfigure[\label{subfig:grid_airfoil}]{
                \includegraphics[width=0.31\textwidth]{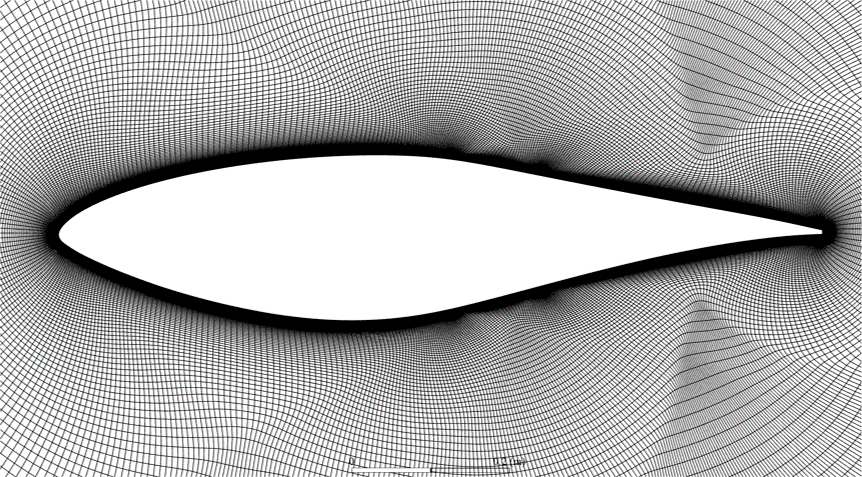}}
                \subfigure[\label{subfig:truncated_TE}]{
                \includegraphics[width=0.31\textwidth]{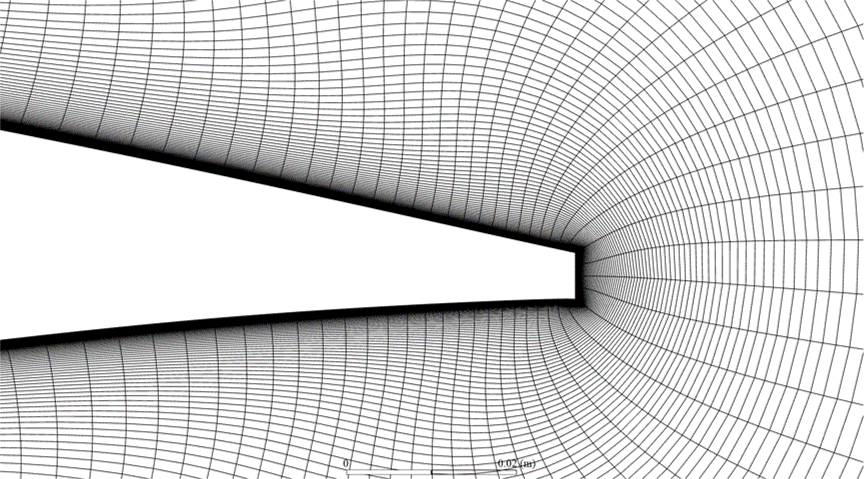}}
                \caption{(a) Structured \textit{O}-grid, (b) mesh around S809 airfoil and (c) detail around its trailing edge.}
                \label{fig:mesh}
            \end{figure*}

The computational domain was discretized through an O-shaped grid, as shown in figure \ref{subfig:O-grid}.
A detailed view of the mesh in the airfoil region is shown in figure \ref{subfig:grid_airfoil}.
To ease the meshing, the airfoil trailing edge was artificially truncated at 97\% of the original chord (see figure \ref{subfig:truncated_TE}).
As shown in section \ref{sec:results}, this did not compromise significantly the comparison of the simulations' results with measurements.
The definition of the domain size and grid refinement level along the airfoil surface was based on a convergence study discussed in section \ref{sec:validation}.
In the airfoil-normal direction a grid growth ration of 1.10 from the surface towards the far field was imposed and a the number of divisions was chosen to guarantee a maximum $y^+$ of the cells adjacent to the airfoil lower than 3.

\section{Results} \label{sec:results}
\subsection{Assessment of CFD model} \label{sec:validation}

\begin{figure*}
                \centering
                \subfigure[]{
                \includegraphics[width=0.45\textwidth]{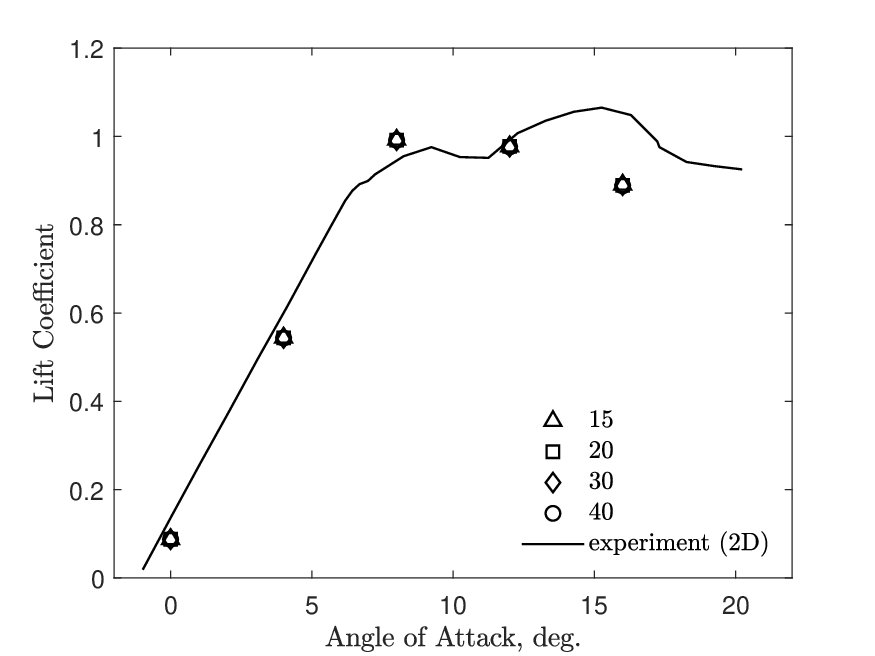}}
                \subfigure[]{
                \includegraphics[width=0.45\textwidth]{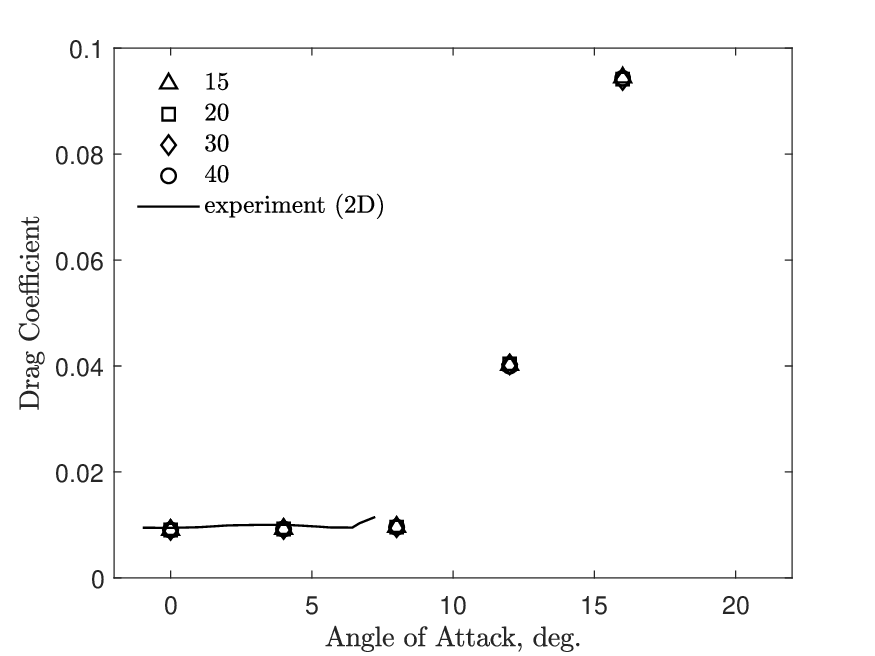}}
                \caption{(a) Lift and (b) drag coefficients. Computations for different domain radii in meters (markers) and experimental data from \citet{somers:1997} (solid lines).}
                \label{fig:radial-domain}
            \end{figure*}

Figure \ref{fig:radial-domain} presents the sensitiveness of the predicted aerodynamic coefficients to the domain radial size, together with wind-tunnel measurements \citep{somers:1997}.
The results were obtained from a series of simulations for conditions without rotational effects considering Reynolds number of 1 million, based on the chord and free-stream velocity. For these flow conditions, there was not appreciable variation of the solution in relation to domain radial size. For angles of attack up to 12 degrees, it is also noticed a good correlation with respect to experimental data. Radius equal to 30 meters was chosen in order to prevent influence of the outer boundary at rotational conditions with increased lift.

\begin{figure*}
                \centering
                \subfigure[]{
                \includegraphics[width=0.45\textwidth]{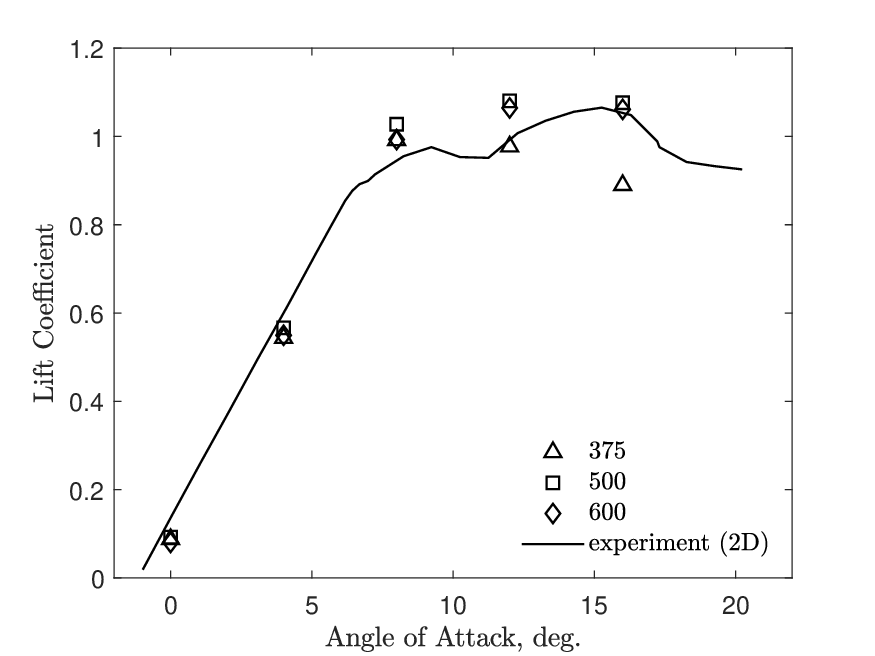}}
                \subfigure[]{
                \includegraphics[width=0.45\textwidth]{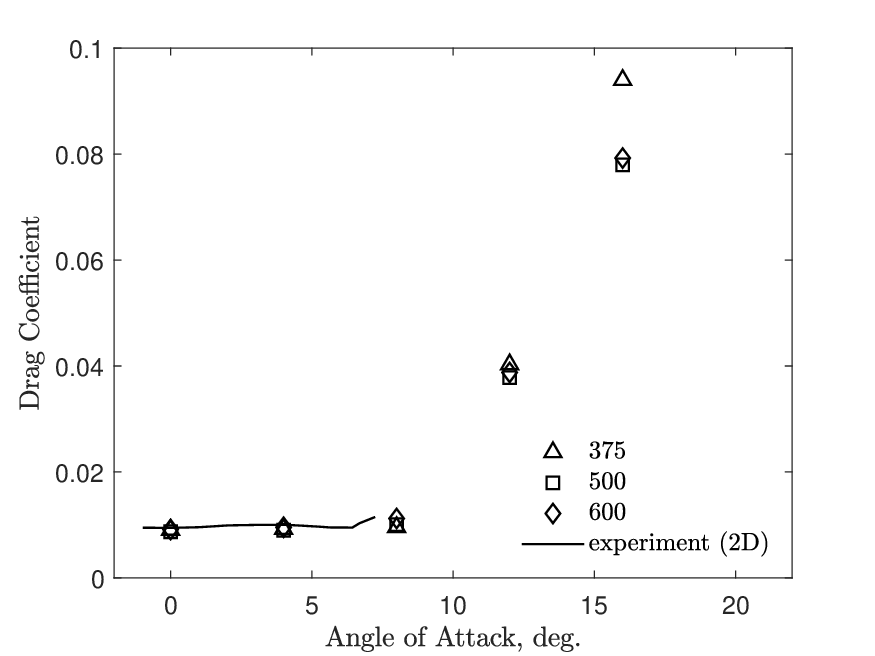}}
                \caption{(a) Lift and (b) drag coefficients. Computations for different number of points along the airfoil surface (markers) and experimental data from \citet{somers:1997} (solid lines).}
                \label{fig:npoints-domain}
            \end{figure*}

Figure \ref{fig:npoints-domain} shows predictions for meshes with different numbers of points along the airfoil.
It was observed that the solution did not vary significantly for meshes exceeding 500 points along the airfoil surface. It is important to stress that divergences between simulations and experimental data for high angles of attack (see Figure \ref{fig:radial-domain}) were then associated to grid refinement.
          
\begin{figure}
    \centering
    \includegraphics[width=\columnwidth]{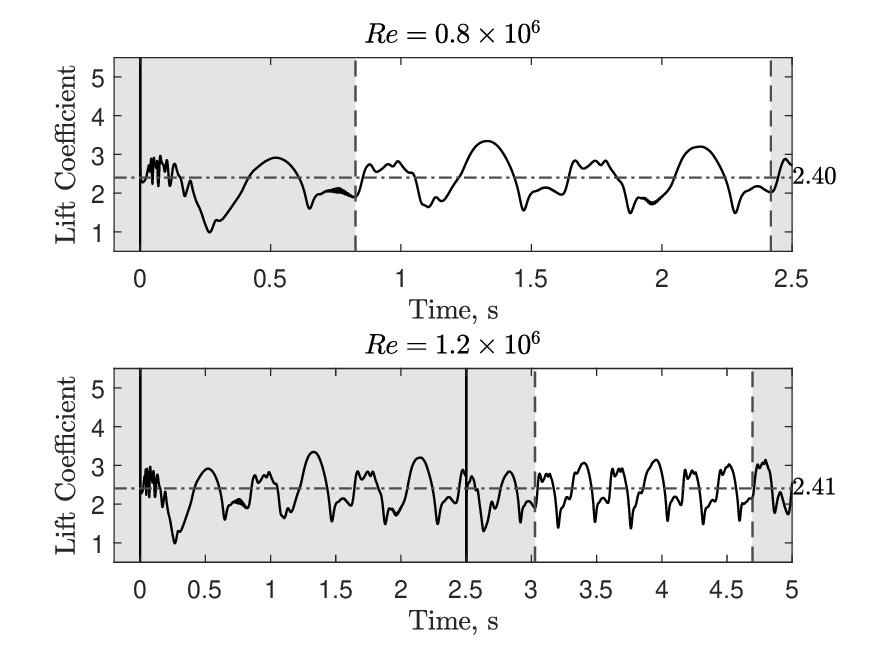}
    \caption{Temporal variations (solid lines) and mean values (horizontal dash-dotted lines) of the lift coefficients for $\alpha$=45 deg (case without rotation) considering different Reynolds numbers.}
    \label{fig:ClCd_Re}
\end{figure}

Figure \ref{fig:ClCd_Re} illustrates the temporal variations of the lift coefficients for Reynolds numbers equal to 0.8$\times$10$^6$ and 1.2$\times$10$^6$ considering the two-dimensional case for the angle of attack of 45 degrees.
The statistical means obtained for each case were evaluated between the vertical dashed lines.
It might be inferred that the flow field with $Re$=1.2$\times$10$^6$ had higher lift coefficient oscillation frequency, although its temporal average did not change significantly. Based on 3D simulations of a small-scale HAWT, \citet{mauro2:2017} suggested a limited influence of Reynolds number for aerodynamics of blade sections within this range of values. Thus, $Re$=1$\times$10$^6$ was chosen as representative condition and the other cases considered here were then simulated at this Reynolds number.

\begin{figure}
                \centering
                \includegraphics[width=\columnwidth]{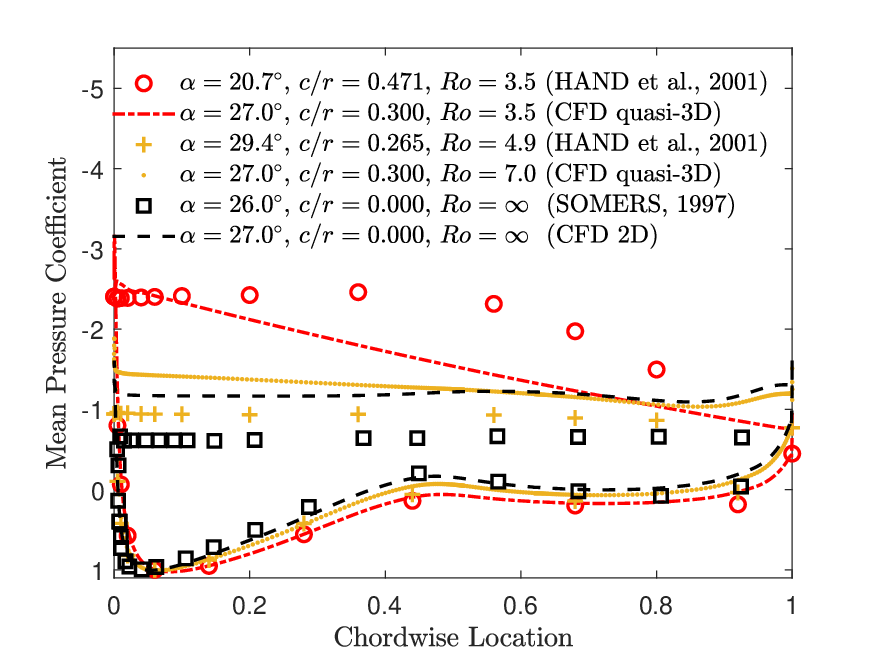}
                \caption{Mean pressure coefficient distributions for $\alpha \approx$ 27 deg.}
                \label{fig:mean-cp}
            \end{figure}

Figure \ref{fig:mean-cp} shows the mean pressure coefficient ($C_p$) distribution on the airfoil surface obtained for simulations with and without rotation in comparison with experimental data \citep{hand:2001,somers:1997}.

It is noted that the simulations captured qualitatively well the rotational effect on the $C_p$ distribution, i.e. the increase in suction on the upper surface.
However, there were some differences in the shape of the distribution in relation to the experimental observations.
It is important to note that the simulations were two-dimensional and based on the RANS approach.
These simplifications are expected to have a significant influence on the dynamics of the separate flow.
For example, the excessive dissipation attributed to the RANS modeling tends to stabilize the flow on the upper surface at high angles of attack, enhancing the time-averaged suction. On the other hand, the absence of three-dimensional structures may result in stronger predicted vortices in separate flows \citep{choudhari:2007}, imposing the same effect on the surface pressure distribution as the RANS' over diffusion.
Note that, for the case without rotation, the upper surface pressure distribution was underestimated, i.e. predicted a stronger suction.
Nonetheless, we argue that the analysis of the simulations can provide insights into relevant aspects of the rotational effects despite the deficiencies mentioned above.

\subsection{Rotational effect on Aerodynamics}

A total of 24 CFD simulations were performed for angles of attack of 10, 15, 27 and 45 degrees. For each angle of attack, flow conditions without and with rotation were considered. As pointed in topic \ref{subsection:quasi-3D-simulations}, values of $c/r$ and $Ro$ of the cases with rotation were chosen to be representative of sections of a small-scale wind turbine. 

\begin{figure*}
                \centering
                \subfigure[]{
                \includegraphics[width=0.45\textwidth]{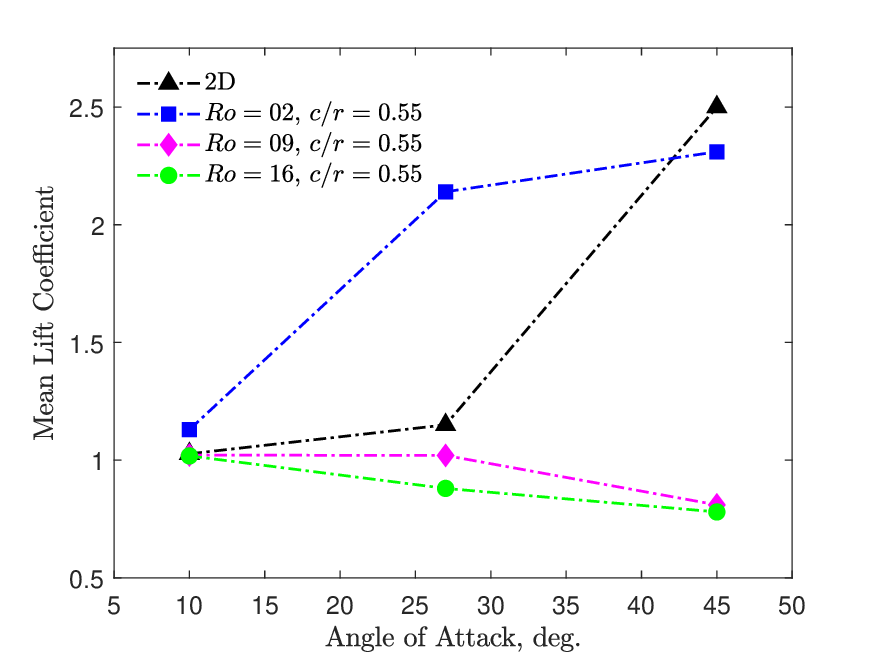}}
                \subfigure[]{
                \includegraphics[width=0.45\textwidth]{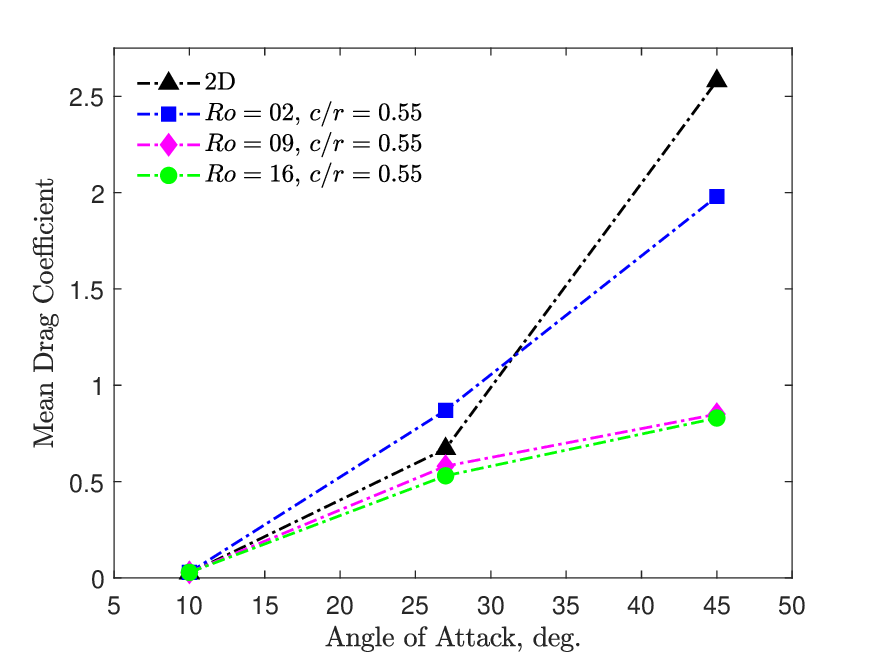}}
                \caption{(a) Lift and (b) drag coefficients as function of angle of attack considering $c/r$ = 0.55 for different Rossby numbers together with 2D case.}
                \label{fig:coef_quasi-3D}
            \end{figure*}

\begin{figure*}
                \centering
                \subfigure[]{
                \includegraphics[width=0.45\textwidth]{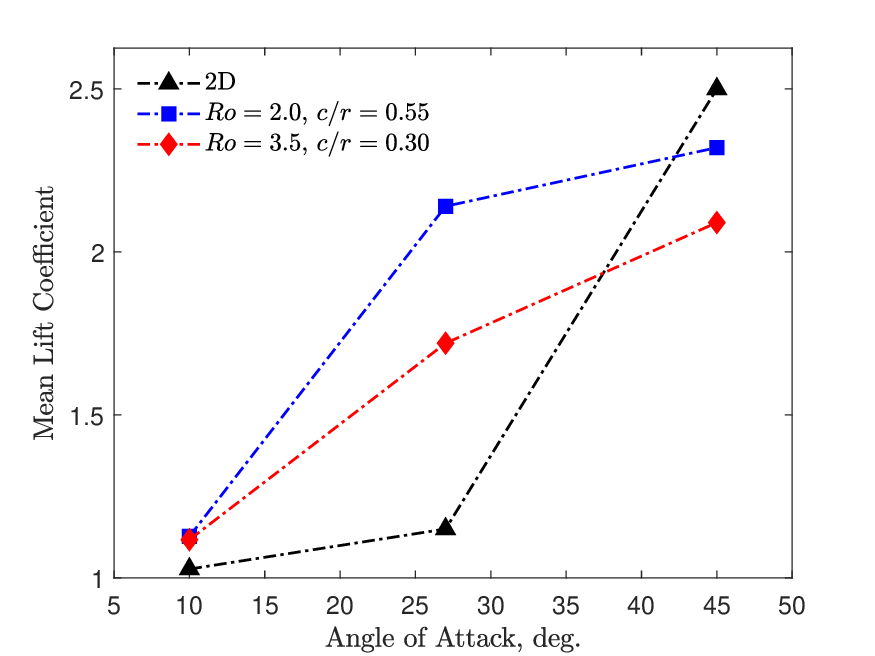}}
                \subfigure[]{
                \includegraphics[width=0.45\textwidth]{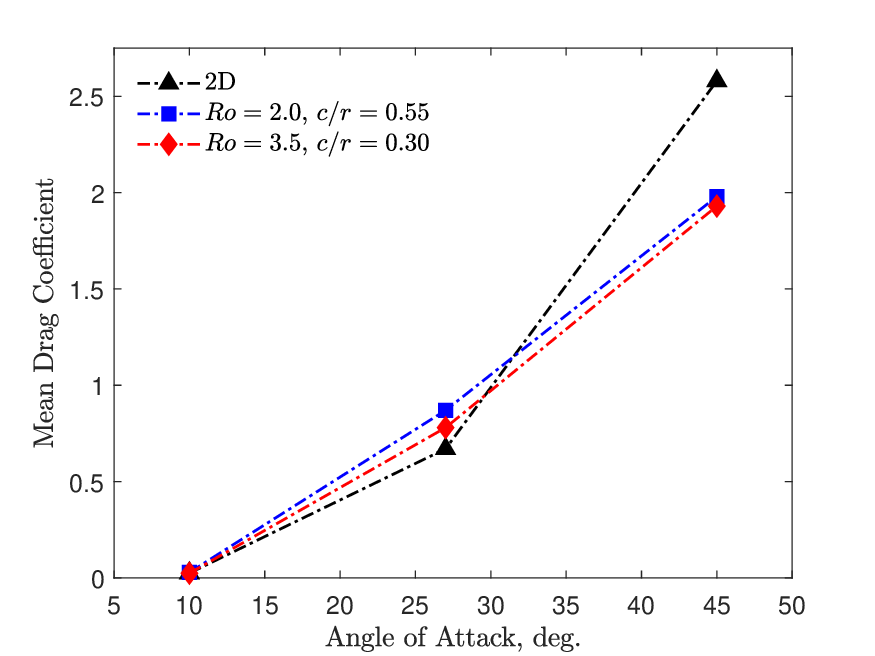}}
                \caption{(a) Lift and (b) drag coefficients as function of angle of attack considering different chord-to-radius ratios for similar Rossby numbers together with 2D case.}
                \label{fig:coef_quasi-3D_c-r}
            \end{figure*}

Table \ref{tab:simulated-cases} denotes the set of cases chosen for the present analysis as well as the lift and drag coefficients predicted by the simulations. For the conditions with rotation, the columns for time-averaged coefficients of lift ($\bar{C_l}$) and drag ($\bar{C_d}$) denote the difference to the non-rotating case at the same angle of attack. Moreover, Figure \ref{fig:coef_quasi-3D} compares cases selected from Table \ref{tab:simulated-cases} with fixed $c/r$ and different Rossby numbers, and Figure \ref{fig:coef_quasi-3D_c-r} compares cases with similar $Ro$ and different chord-to-radius ratios.

It is seen that the rotational effects on forces were stronger for high angles of attack. Interestingly, the lift augmentation was significant only for sufficiently low values of $Ro$ and $c/r$ had virtually no influence on the drag coefficient. Moreover, it was observed that, whether rotation increases or reduces  $\bar{C_d}$, depends on the angle of attack. Surprisingly, for an angle of attack equal to 45 deg, the rotation decreased $\bar{C_l}$.

\begin{table}[H]
                \centering
                \caption{Set of simulated cases. For cases with rotation, the columns for time-averaged coefficients of lift ($\bar{C_l}$) and drag ($\bar{C_d}$) denote the difference to the respective non-rotating case (boldfaced). Asterisks indicate cases chosen for further analysis in the next topics.}
                \begin{ruledtabular}
                \begin{tabular}{clcll}
                     \multicolumn{1}{c}{$\alpha$ (deg)} & \multicolumn{1}{c}{$c/r$} & \multicolumn{1}{c}{$Ro$} & \multicolumn{1}{c}{$\bar{C_l}$} & \multicolumn{1}{c}{$\bar{C_d}$} \\
                    \hline
                    \textbf{10} & \textbf{0} & $\boldsymbol{\infty}$ & \textbf{ \;1.03} & \textbf{ \;0.0245}* \\
                    10 & 0.0625 & 16 & +0.02 &  \:-0.0009* \\
                    10	& 0.3 & 3.5 & +0.09 &  \:-0.0004 \\
                    10	& 0.3 & 7 & +0.04 &  \:-0.0003 \\
                    10 & 0.55 & 2 & +0.10 & +0.0039 \\
                    10	& 0.55 & 7 &  \:-0.01 & +0.0038 \\
                    10	& 0.55 & 9 &  \:-0.01 & +0.0039 \\
                    10	& 0.55 & 16 &  \:-0.01 & +0.0040 \\
                    \textbf{15} & \textbf{0} & $\boldsymbol{\infty}$ & \textbf{ \;1.20} & \textbf{ \;0.0555}* \\
                    15 & 0.45 & 3 & +0.30 & +0.0016* \\ 
                    \textbf{27} & \textbf{0} & $\boldsymbol{\infty}$ & \textbf{ \;1.15} & \textbf{ \;0.67}* \\
                    27	& 0.0625 & 16 & +0.08 & +0.02 \\
                    27 & 0.3 & 3.5 & +0.57 & +0.11* \\
                    27	& 0.3 & 7 & +0.13 & +0.02 \\
                    27	& 0.55 & 2 & +0.99 & +0.20* \\
                    27	& 0.55 & 9 &  \:-0.13	&  \:-0.09* \\
                    27	& 0.55 & 16 &  \:-0.27 &  \:-0.14* \\
                    \textbf{45} & \textbf{0} & $\boldsymbol{\infty}$ & \textbf{ \;2.50} & \textbf{ \;2.58} \\
                    45	& 0.0625 & 16 & \:-0.13 &  \:-0.05 \\
                    45	& 0.3 & 3.5 & \:-0.41 &  \:-0.65 \\
                    45	& 0.3 & 7 &  \:-0.94 &  \:-1.00 \\
                    45	& 0.55 & 2 &  \:-0.18 &  \:-0.60 \\
                    45	& 0.55 & 9 &  \:-1.69 &  \:-1.73 \\
                    45	& 0.55 & 16 &  \:-1.72 &  \:-1.75 \\
                \end{tabular}
                \end{ruledtabular}
                \label{tab:simulated-cases}
            \end{table}

\begin{figure*}
                \centering
                \subfigure[]{
                \includegraphics[width=0.45\textwidth]{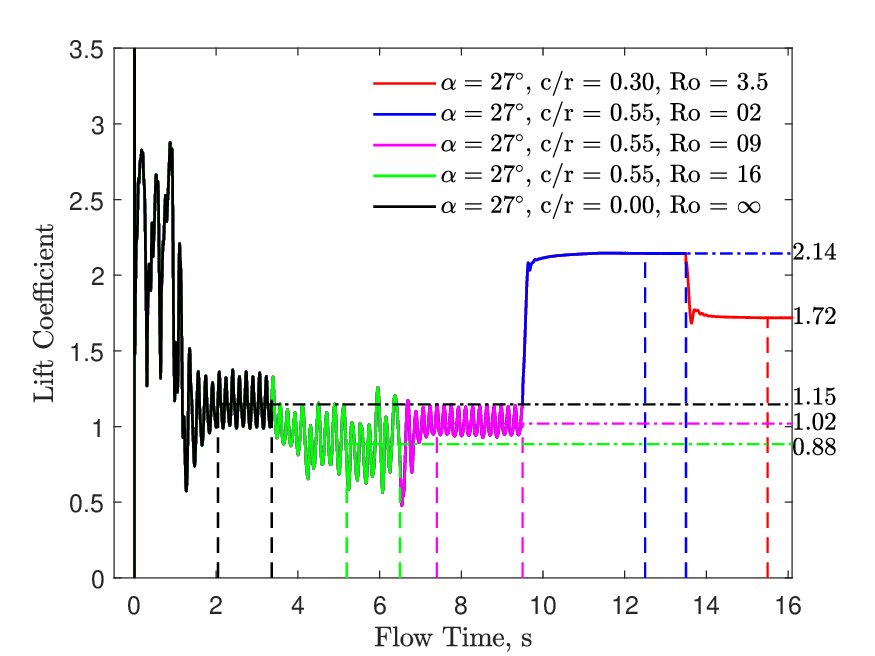}}
                \subfigure[]{
                \includegraphics[width=0.45\textwidth]{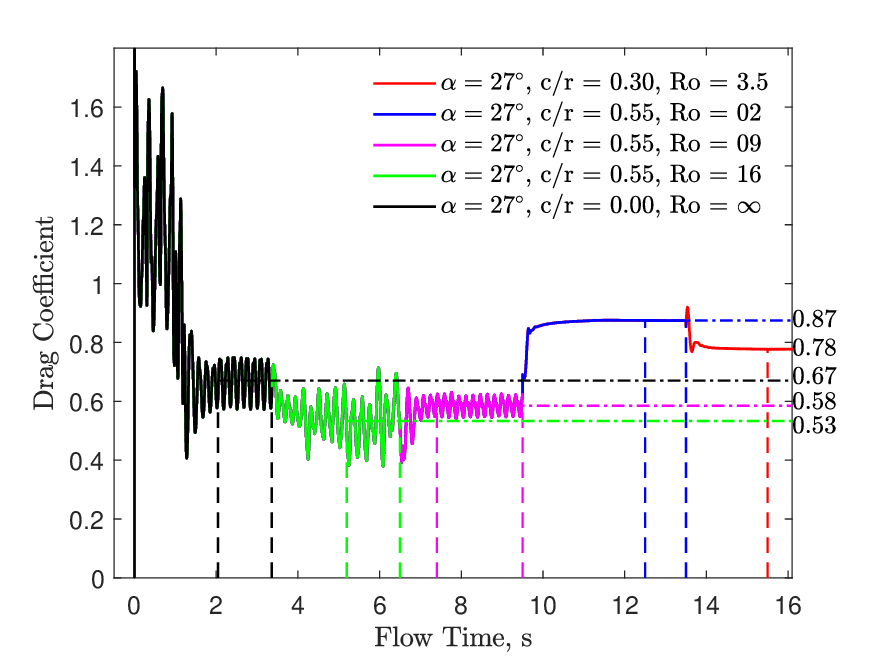}}
                \caption{Temporal evolutions (solid lines) and statistical averages (horizontal dash-dotted lines) of the (a) lift and (b) drag coefficients for $\alpha$ = 27 deg and $c/r$ = 0.55 considering different Rossby numbers together with 2D case.}
                \label{fig:ClCd_alpha_27}
            \end{figure*}

Figure \ref{fig:ClCd_alpha_27} shows temporal evolution and average values of the lift and drag coefficients for an angle of attack of 27 degrees, $c/r$ equal to 0.55 and different Rossby numbers along with non-rotating case. Values indicated on the right of the graphs refer either to the temporal average of the referred coefficient or to the converged value for steady-state solutions, whereas vertical dashed lines indicate the intervals considered for taking temporal average of oscillatory solutions.

For conditions without rotation and for high $Ro$, simulations presented oscillatory solutions, while for low $Ro$, steady-state solutions were obtained.
Since the Rossby number represents the ratio between the inertial and Coriolis forces, Figure \ref{fig:ClCd_alpha_27} suggests that the Coriolis acceleration had a stabilizing effect on the flow for an angle of attack of 27 deg.

\subsection{Rotational effect on the flow field}
\label{subsection:rotational-effect}

\begin{figure*}
                \centering
                \includegraphics[width=0.45\textwidth]{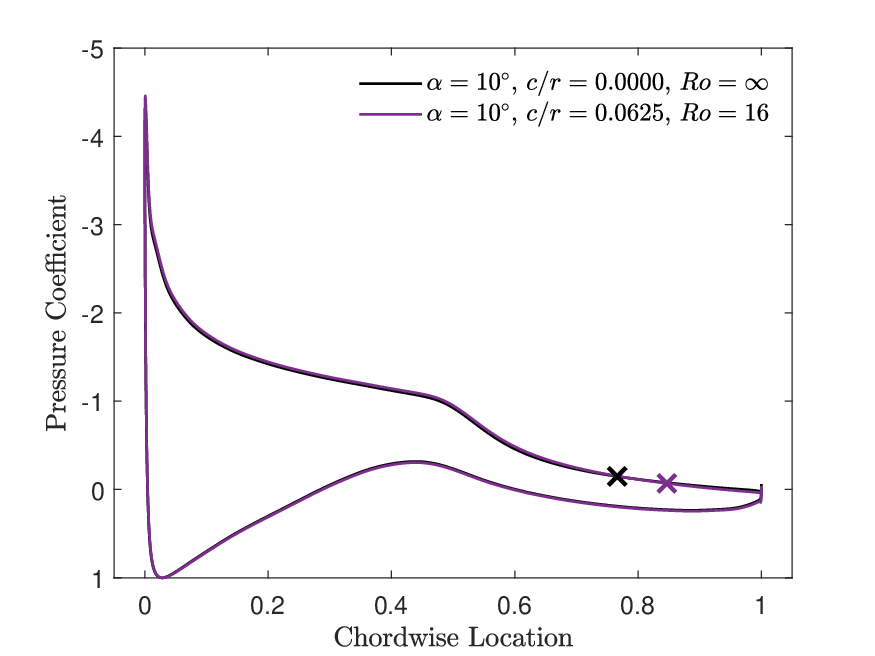}
                \includegraphics[width=0.45\textwidth]{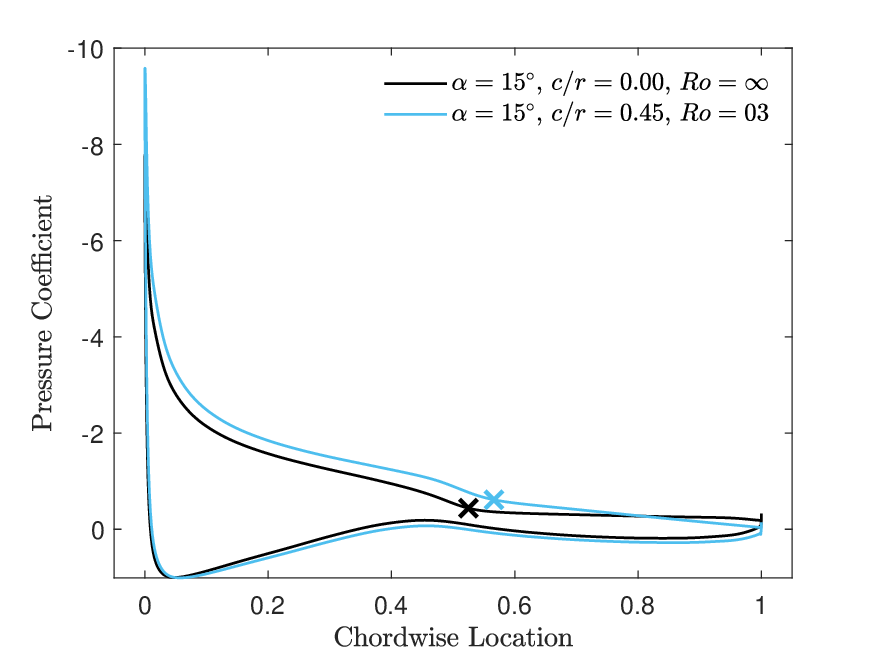}\\
                \includegraphics[width=0.45\textwidth]{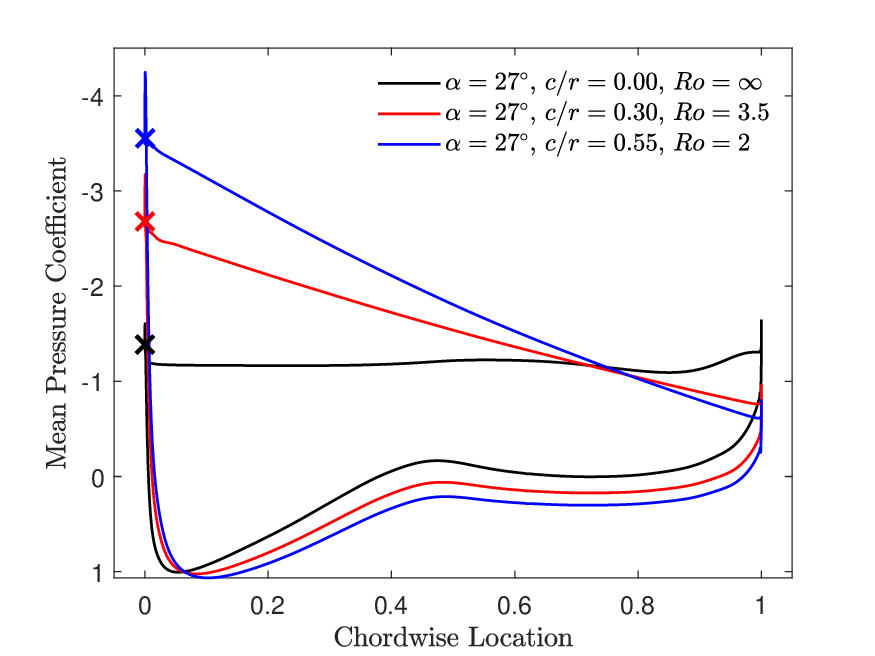}
                \includegraphics[width=0.45\textwidth]{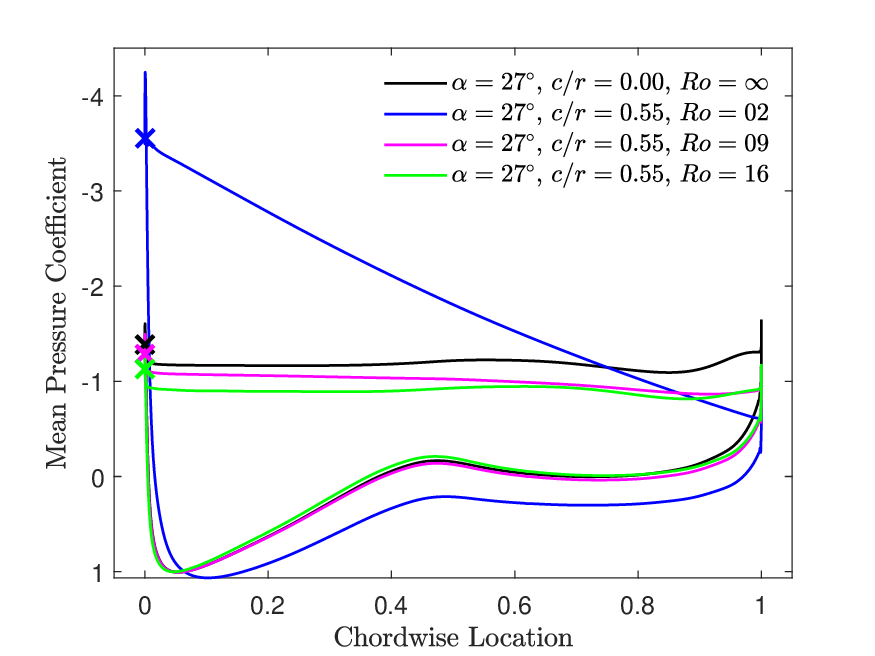}
                \caption{Surface pressure distributions and upper surface separation points (markers) for the cases with and without rotation.}
                \label{fig:Cp_alpha}
            \end{figure*}

The distribution of the pressure coefficient on the airfoil surface together with the position of the boundary-layer separation point on the upper surface for each case is shown in Figure \ref{fig:Cp_alpha}. The separation point was assumed as the locus where the skin frictional coefficient became zero.

Note that for 10 deg and 15 deg angles of attack, a trailing-edge separation was observed, while for 27 deg, the boundary-layer separation occurred at the leading edge. It was also observed that the rotation delayed the separation point for all cases, mainly for the conditions considered at lower angles of attack.
However, this delay did not necessarily have a significant impact on the surface pressure distribution.

It is noteworthy that the simulations captured the upper surface suction enhancement associated with rotation despite the fact that the simulation model cannot capture the centrifugal pumping.
This apparent inconsistency was also observed in span-wise periodic simulations \citep{souza:2020} and suggests that other mechanisms may be at play.

            \begin{figure*}
                \centering
                \subfigure[$\alpha$=10$^{\circ}$; $c/r$=0.0000; $Ro$=$\infty$ (without rotation)]{
                \includegraphics[width=0.34\textwidth]{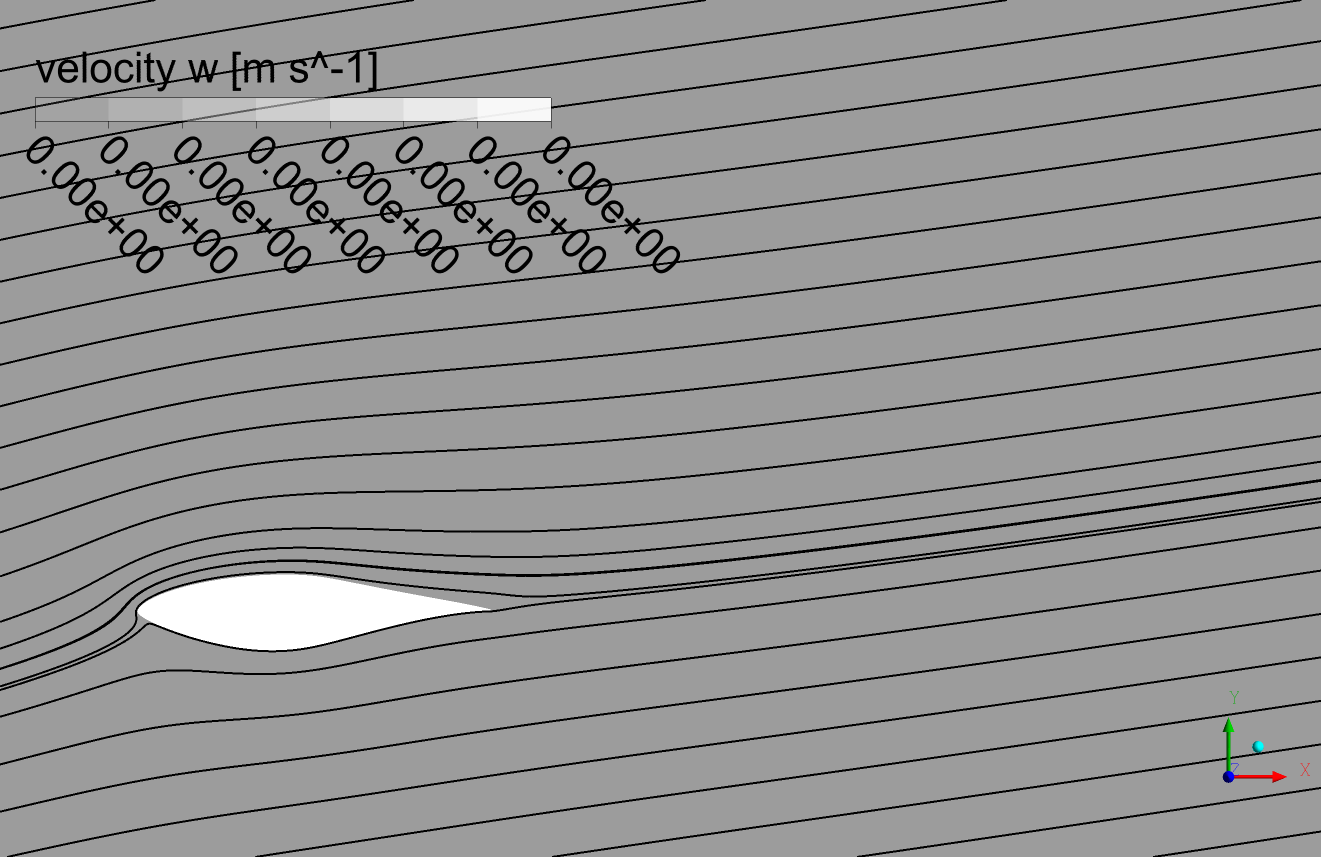}}
                \subfigure[$\alpha$=10$^{\circ}$; $c/r$=0.0625; $Ro$=16 (with rotation)]{
                \includegraphics[width=0.34\textwidth]{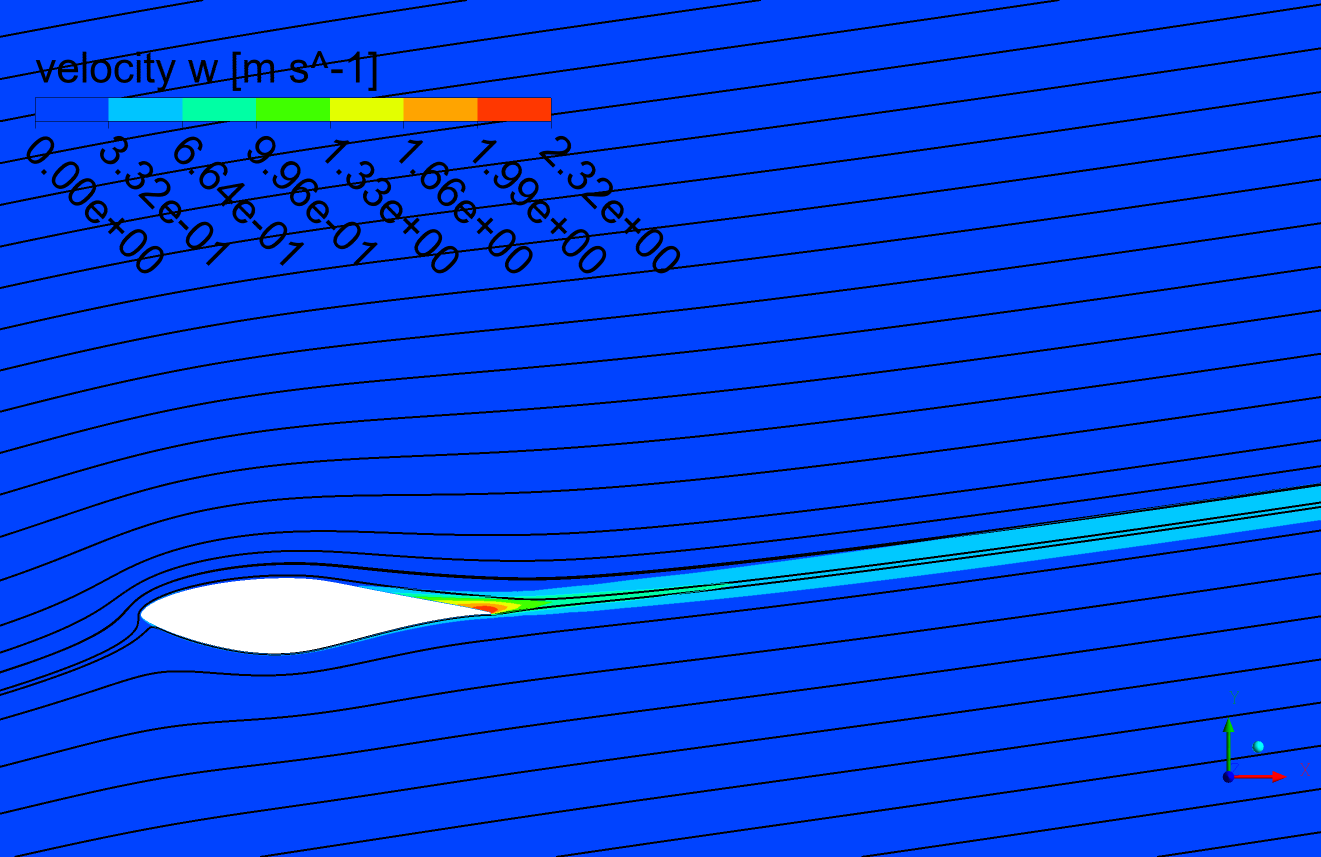}}\\
                \subfigure[$\alpha$=15$^{\circ}$; $c/r$=0.00; $Ro$=$\infty$ (without rotation)]{
                \includegraphics[width=0.34\textwidth]{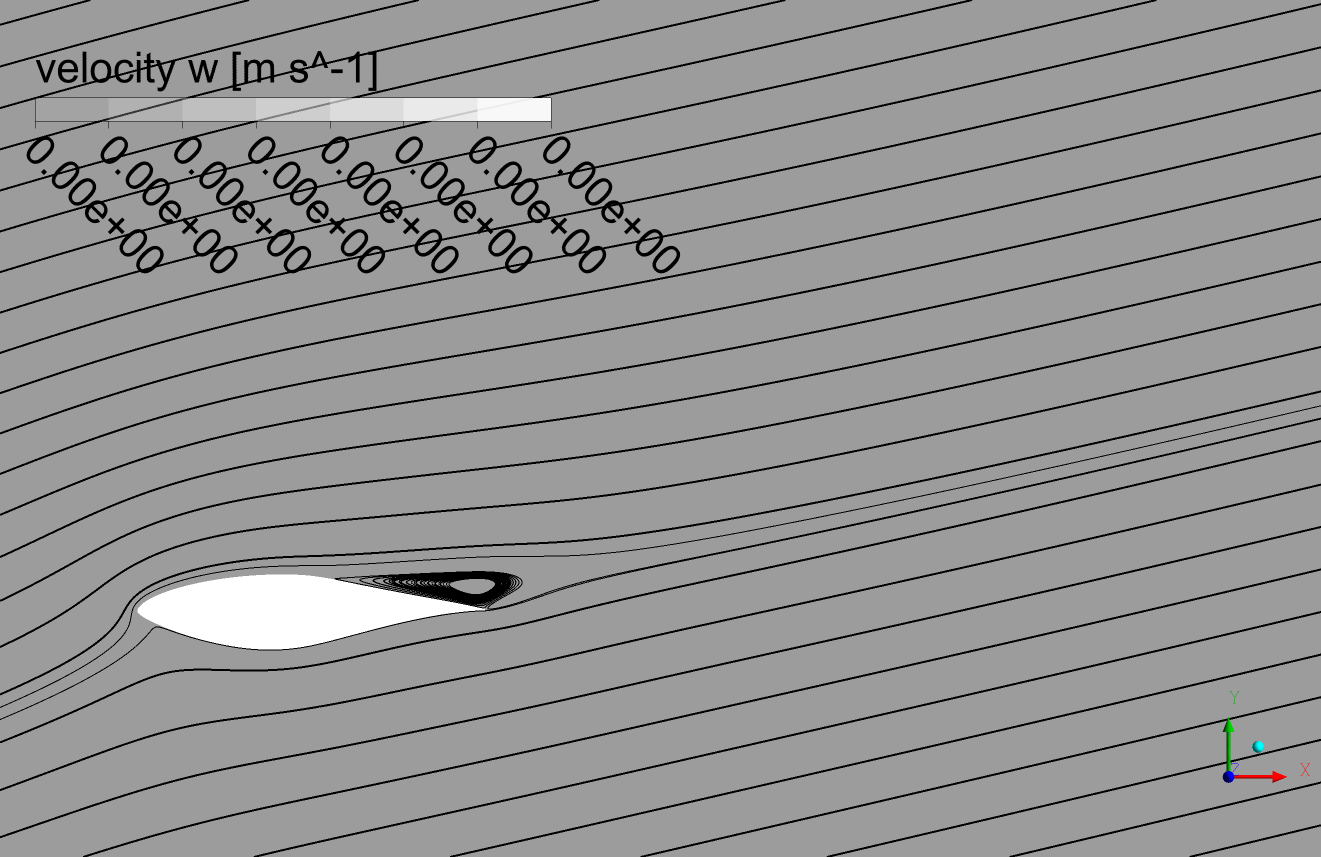}}
                \subfigure[$\alpha$=15$^{\circ}$; $c/r$=0.45; $Ro$=3 (with rotation)]{
                \includegraphics[width=0.34\textwidth]{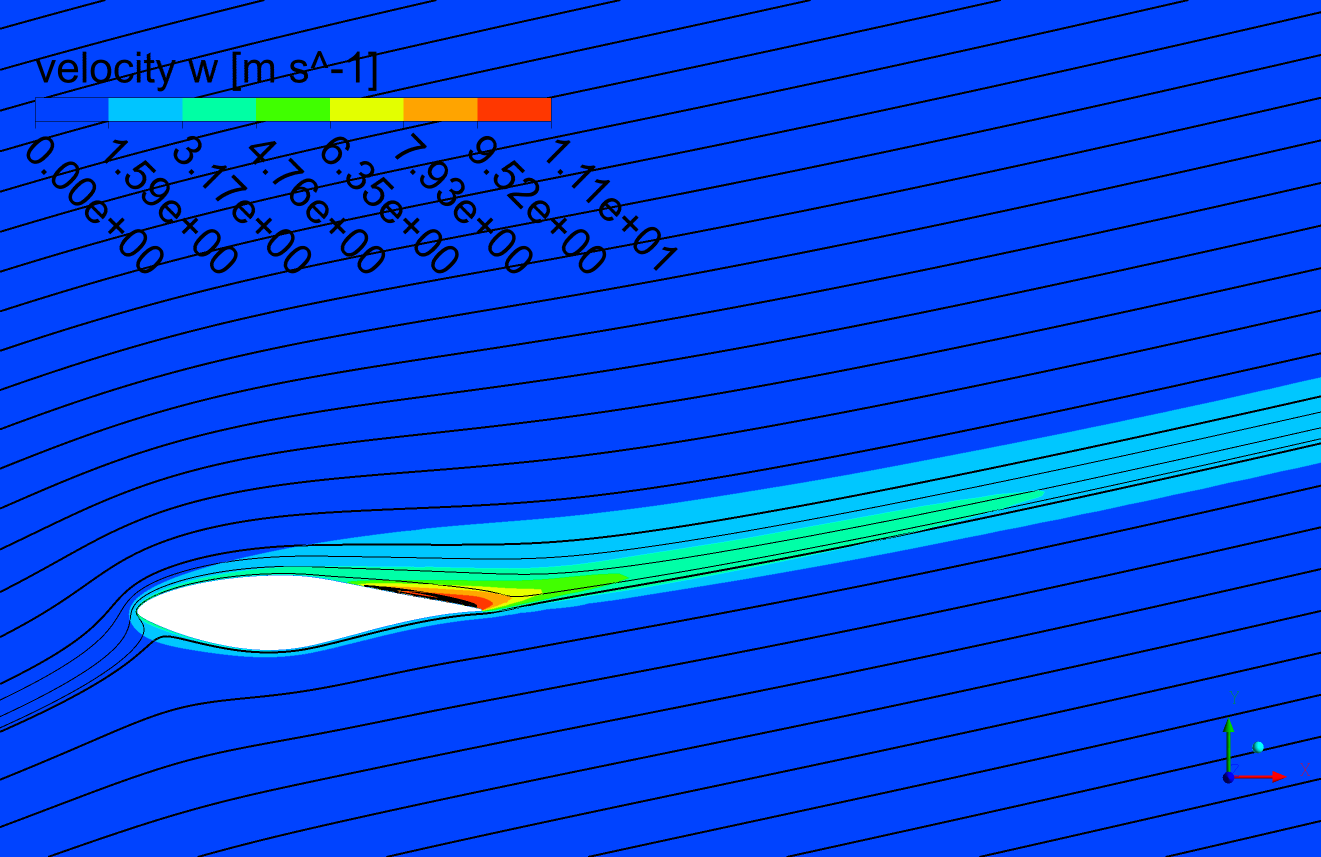}}\\
                \subfigure[$\alpha$=27$^{\circ}$; $c/r$=0.0; $Ro$=$\infty$ (without rotation)]{
                \includegraphics[width=0.31\textwidth]{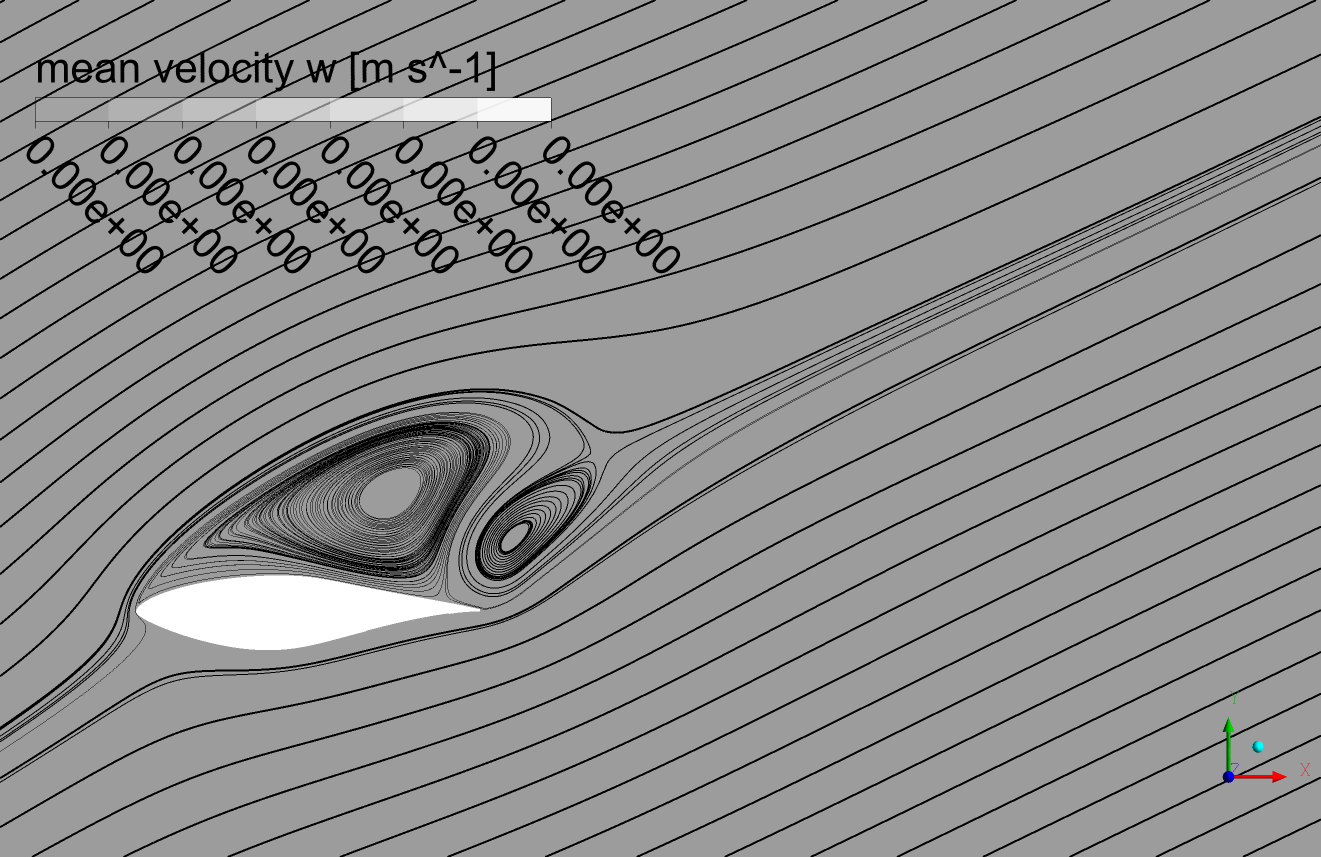}}
                \subfigure[$\alpha$=27$^{\circ}$; $c/r$=0.3; $Ro$=3.5 (with rotation)]{
                \includegraphics[width=0.31\textwidth]{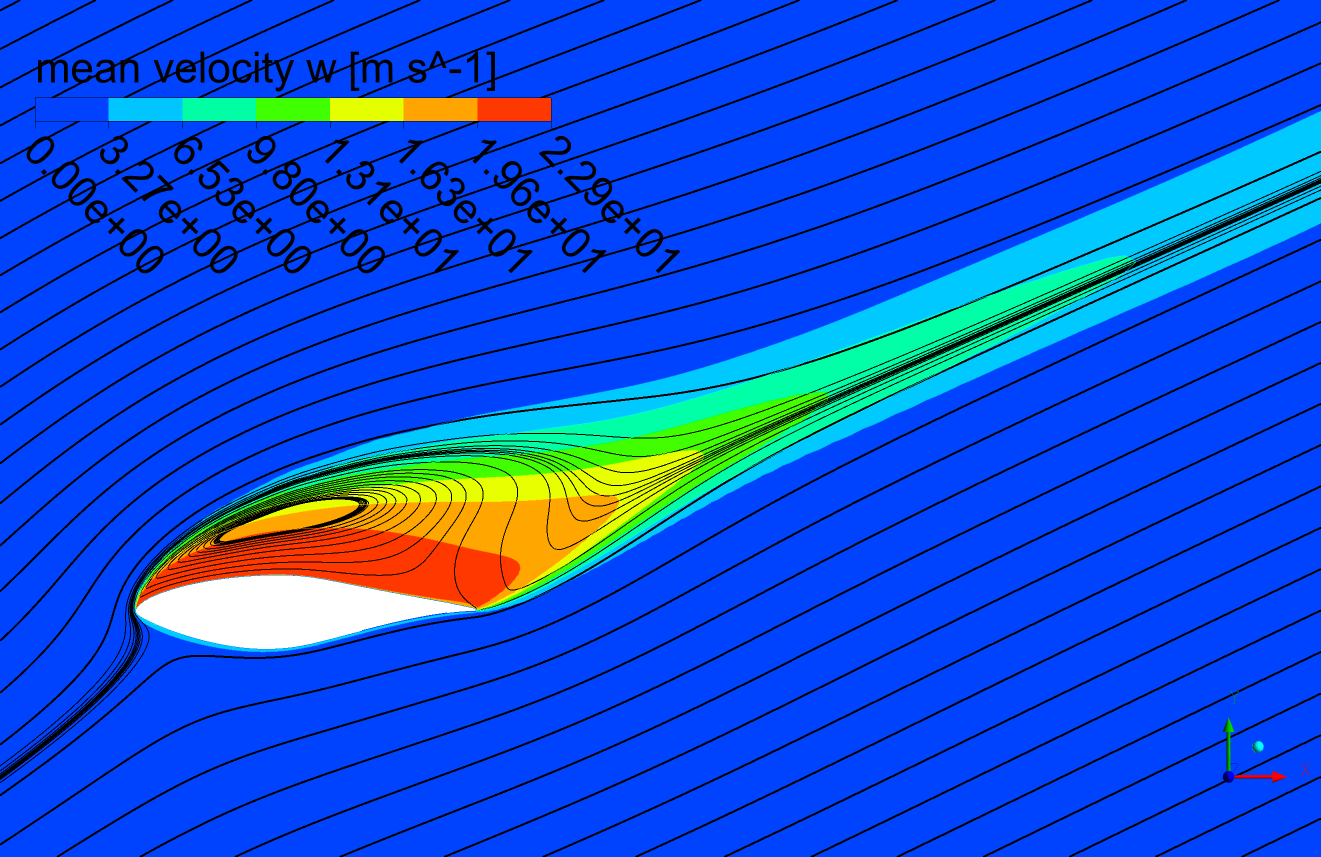}}
                \subfigure[$\alpha$=27$^{\circ}$; $c/r$=0.55; $Ro$=2 (with rotation)]{
                \includegraphics[width=0.31\textwidth]{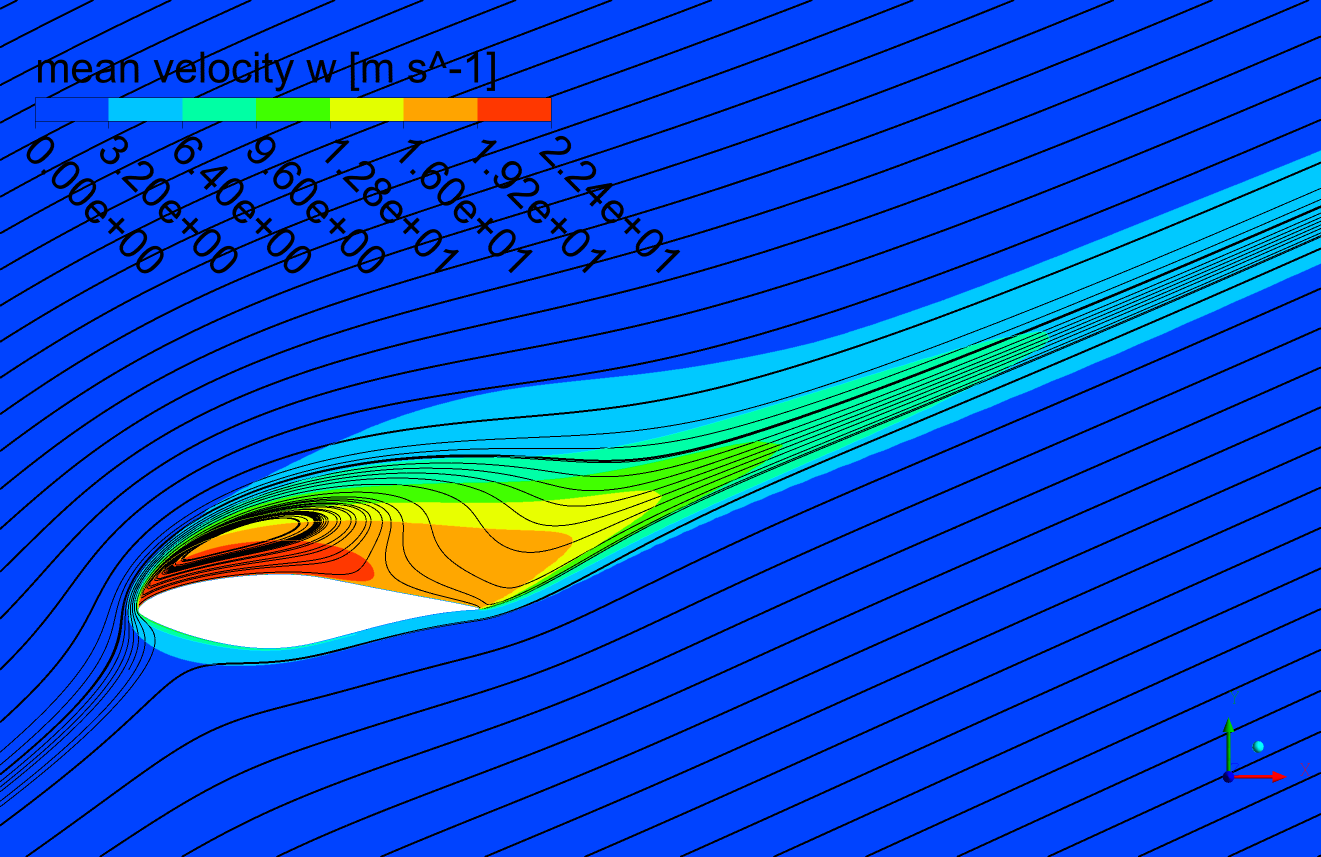}}
                \caption{Field of span-wise velocity along with streamlines for the cases without (left-hand side) and with rotation (center and right-hand side).}
                \label{fig:w_cases}
            \end{figure*}

            \begin{figure*}
                \centering
                \subfigure[$\alpha$ = 27$^{\circ}$; $c/r$ = 0.00; $Ro$ = $\infty$ ($C_{l_{min}}$)]{
                \includegraphics[width=0.45\textwidth]{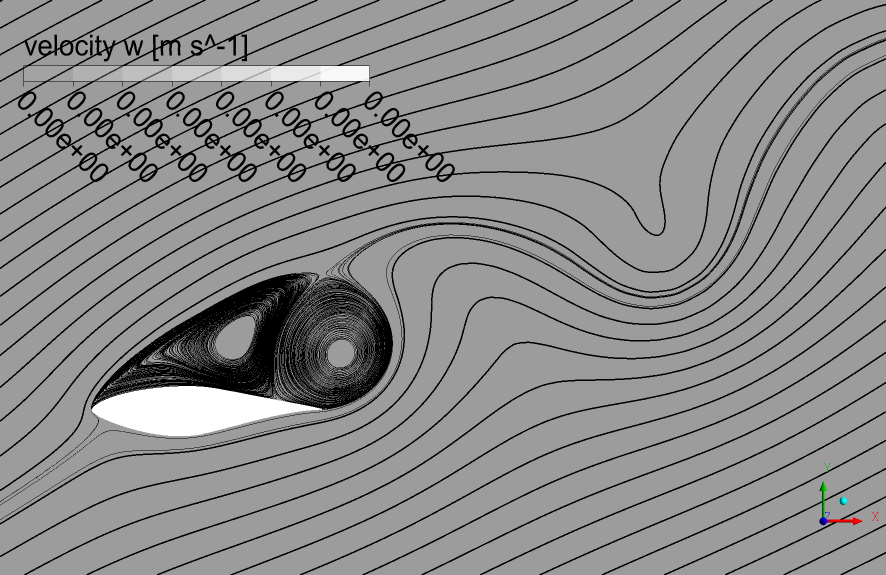}
                \label{subfig:w_27_00_min-a}}
                 \subfigure[$\alpha$ = 27$^{\circ}$; $c/r$ = 0.00; $Ro$ = $\infty$ ($C_{l_{max}}$)]{
                \includegraphics[width=0.45\textwidth]{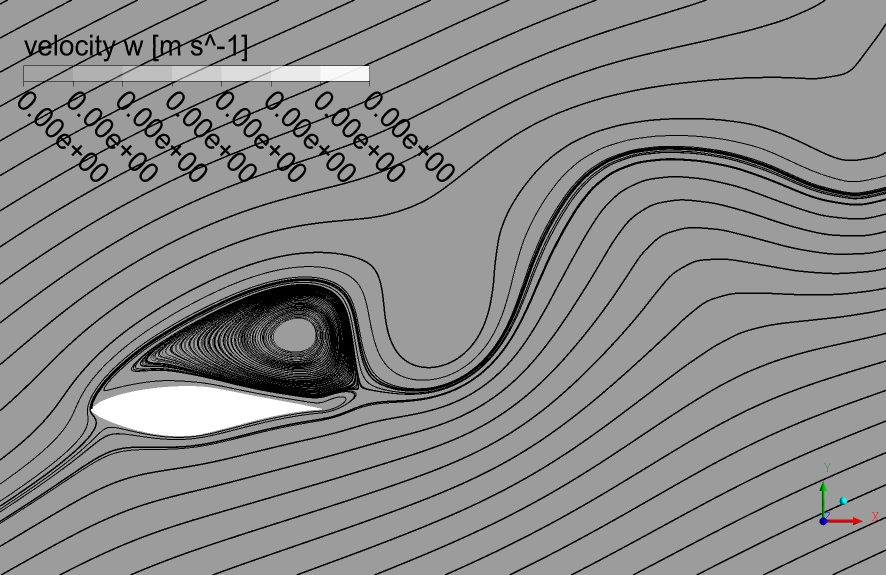}
                \label{subfig:w_27_00_max-b}}
                \subfigure[$\alpha$ = 27$^{\circ}$; $c/r$ = 0.55; $Ro$ = 9 ($C_{l_{min}}$)]{
                \includegraphics[width=0.45\textwidth]{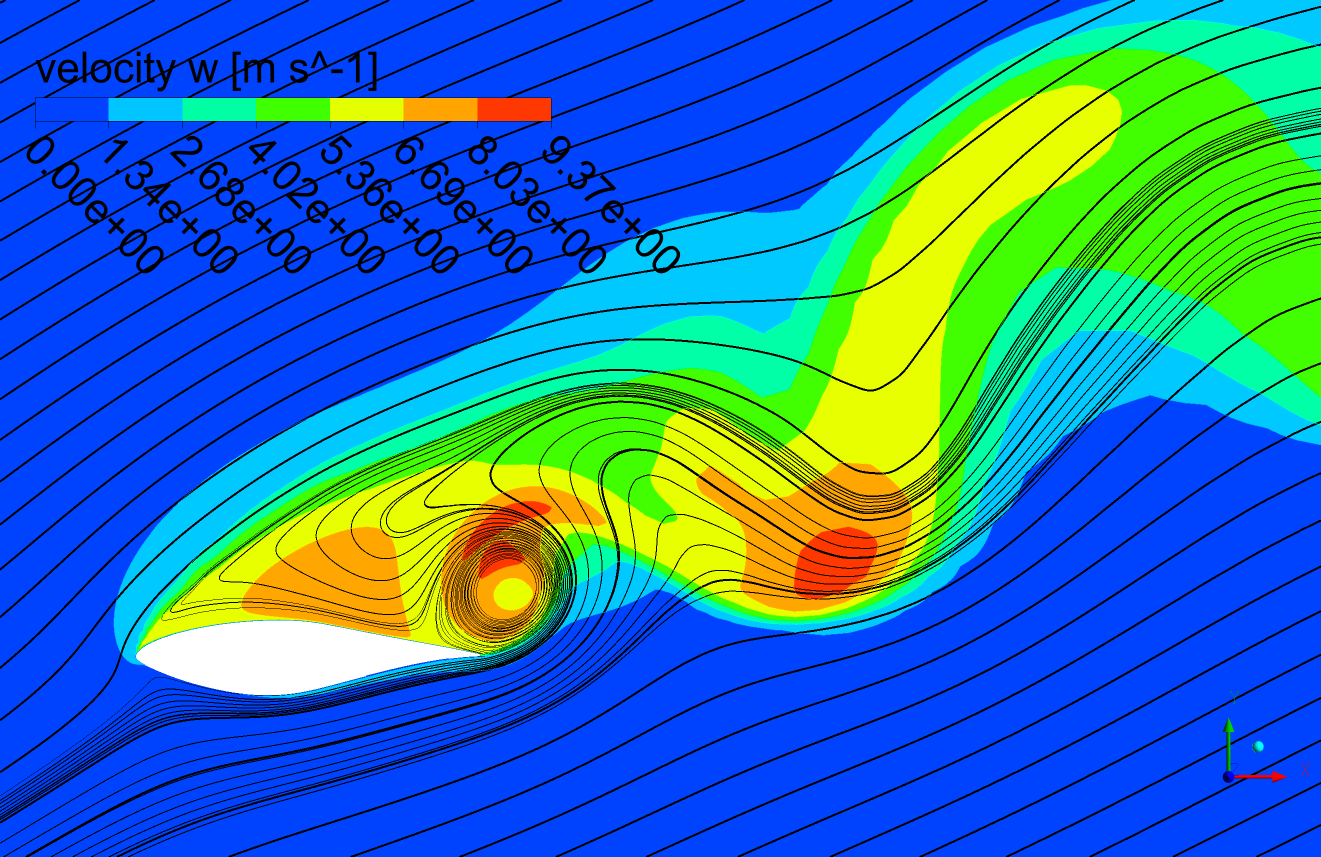}
                \label{subfig:w_27_rot09_min-c}}
                \subfigure[$\alpha$ = 27$^{\circ}$; $c/r$ = 0.55; $Ro$ = 9 ($C_{l_{max}}$)]{
                \includegraphics[width=0.45\textwidth]{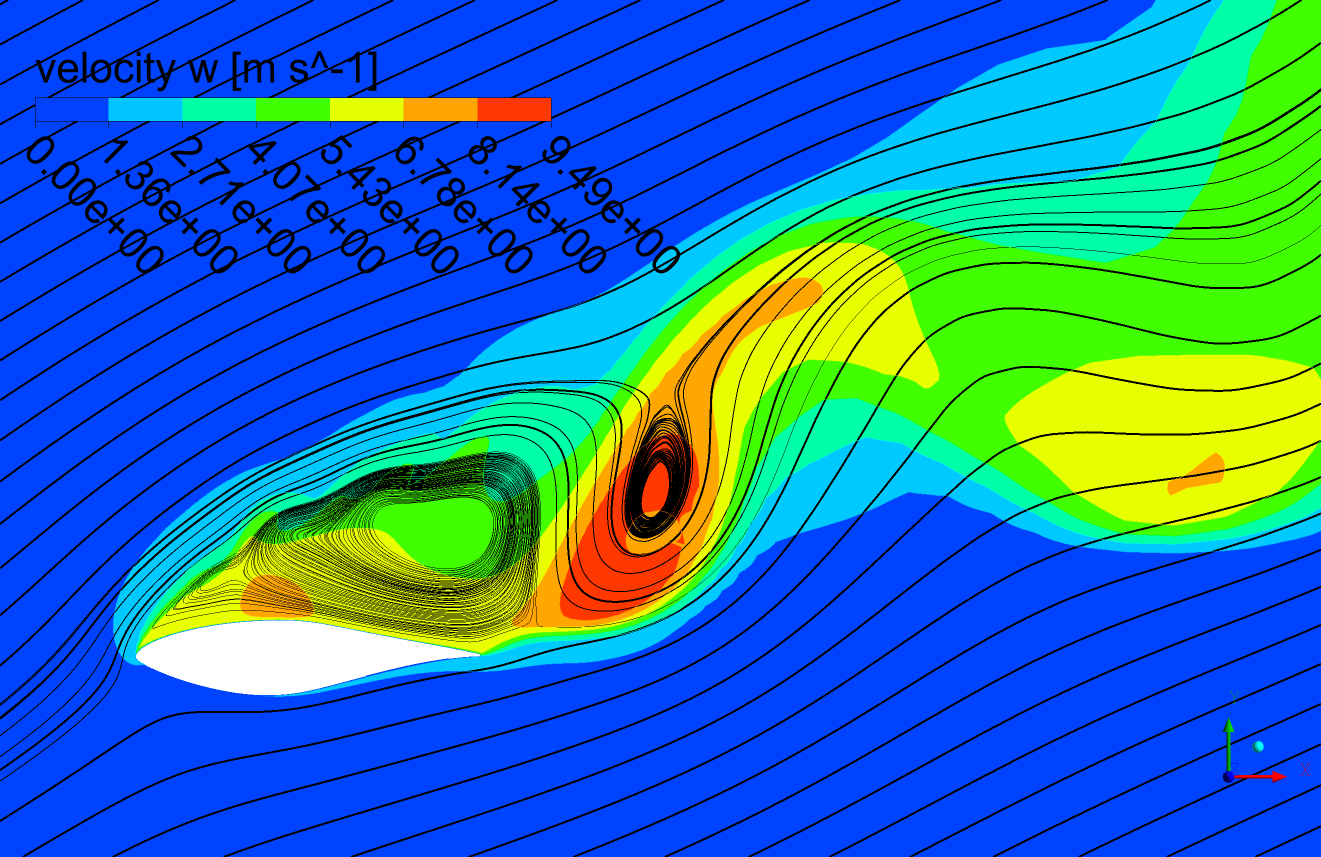}
                \label{subfig:w_27_rot09_max-d}}
                \caption{Field of span-wise velocity for $C_{l_{min}}$ and $C_{l_{max}}$ conditions along with streamlines considering the cases without (top side) and with rotation (bottom side).}
                \label{fig:w_27_Clmin-max}
            \end{figure*}

Figure \ref{fig:w_cases} shows the streamlines and the span-wise velocity component $w$ for angles of attack of 10 deg, 15 deg, for the condition at 27 deg without rotation (based on time-averaged field), and for rotational conditions at 27 deg that converged to a steady-state solution.
It is seen that $w$ was more significant in separate flow region. In this region, the particles move more slowly and, therefore, suffer greater trajectory deflection by the centrifugal force.

For $\alpha$ = 10 deg, the rotation had only a mild effect on the flow topology. For $\alpha$ = 15 deg, the rotation reduced the height of the separate flow region near the trailing edge. In this case, the lift increase may be associated to the weakening of the decambering effect, as explained by \citet{bangga:2017b}.

For $\alpha$ = 27 deg, the flow field had different characteristics in comparison to the two others: the boundary layer separated at the leading edge and there was a reattachment, which on average occurred close to the trailing edge, and moved upstream as $Ro$ reduced and $c/r$ increased. This behavior corresponds to a stationary leading-edge vortex.

Figure \ref{fig:w_27_Clmin-max} shows streamlines and contours of $w$ for $\alpha$ = 27 deg, without rotation, and with Rossby number equal to 9. At these conditions the solution was unsteady. Two instants are shown for each case, one of minimum $C_l$ and one of maximum $C_l$. The figure indicates periodic shedding of the leading-edge vortex alternated by the shedding of the trailing-edge vortex, typical dynamics of flow over bluff bodies or airfoils at high angle of attack.

As mentioned in section \ref{sec:introduction}, analyses of the unsteady pressure measurements from the Phase-VI experiments performed by \citet{schreck:2007} revealed a correlation between high rotational augmentation and the presence of two spots of high pressure root mean square on the section suction surface, which the authors attributed to a standing vortex.

It is recognized that the ``strictly'' steady solution obtained for the cases of figures \ref{fig:w_cases}e through \ref{fig:w_cases}g is an artifact of the overly simplified model. However, in view of the aforementioned experimental observations, the present results suggest that at, an angle of attack of 27 deg, for sufficiently low Rossby number, the rotation may stabilize the leading-edge vortex on the section suction surface.

\subsection{Vorticity tilting}
\label{subsection:vorticity-tilting}

From the sectional momentum equation, equation \ref{eq:xyMomentum}, we can obtain the span-wise vorticity equation and thus the Coriolis contribution to it,

\begin{equation}
    S_{\omega} = - 2 \Omega \left( \frac{\partial w}{\partial x} \sin{\varphi} + \frac{\partial w}{\partial y} \cos{\varphi} \right). \nonumber
    \label{eq:source}
\end{equation}

\begin{figure*}
                \centering
                \subfigure[$\alpha$=10$^{\circ}$; $c/r$=0.0000; $Ro$=$\infty$ (without rotation)]{
                \includegraphics[width=0.34\textwidth]{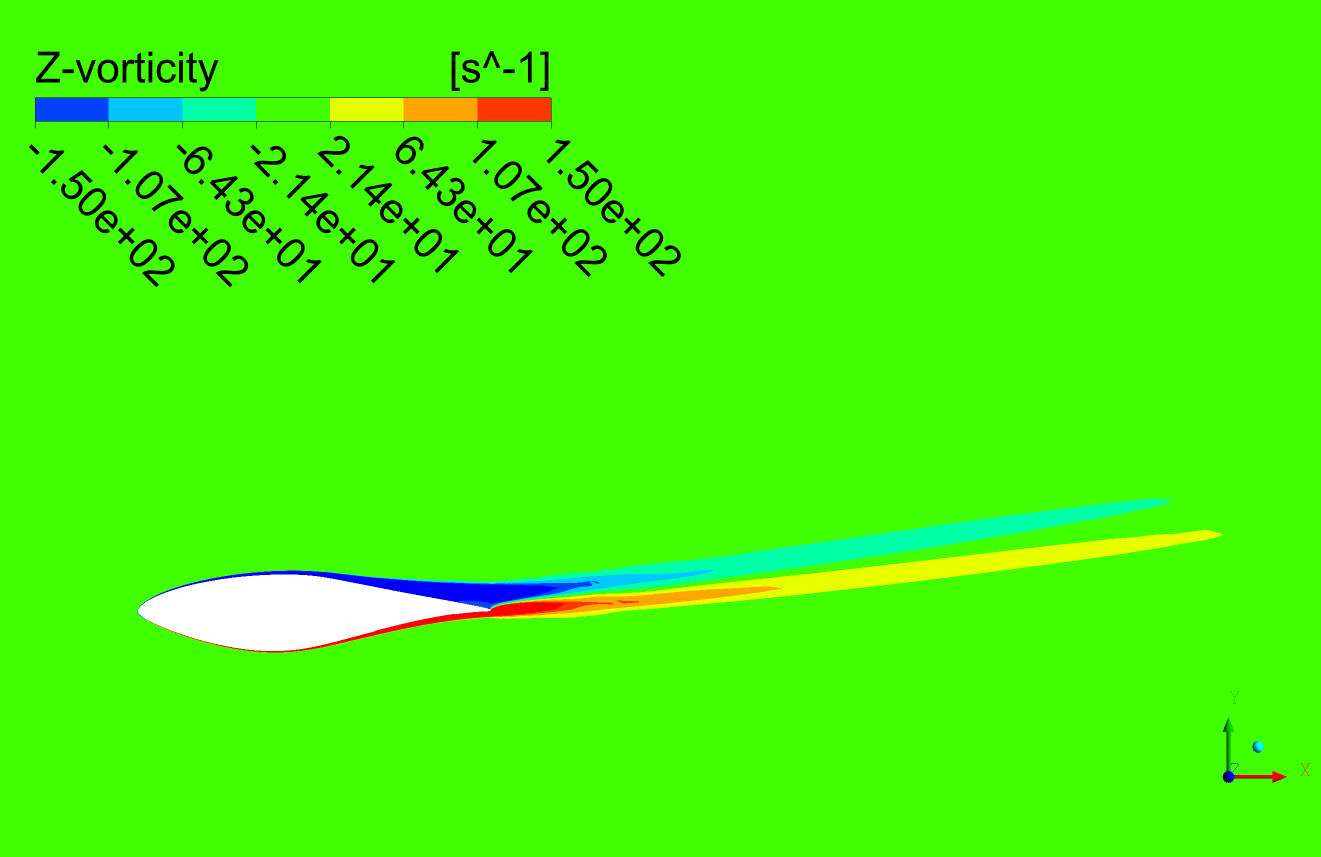}}
                \subfigure[$\alpha$=10$^{\circ}$; $c/r$=0.0625; $Ro$=16 (with rotation)]{
                \includegraphics[width=0.34\textwidth]{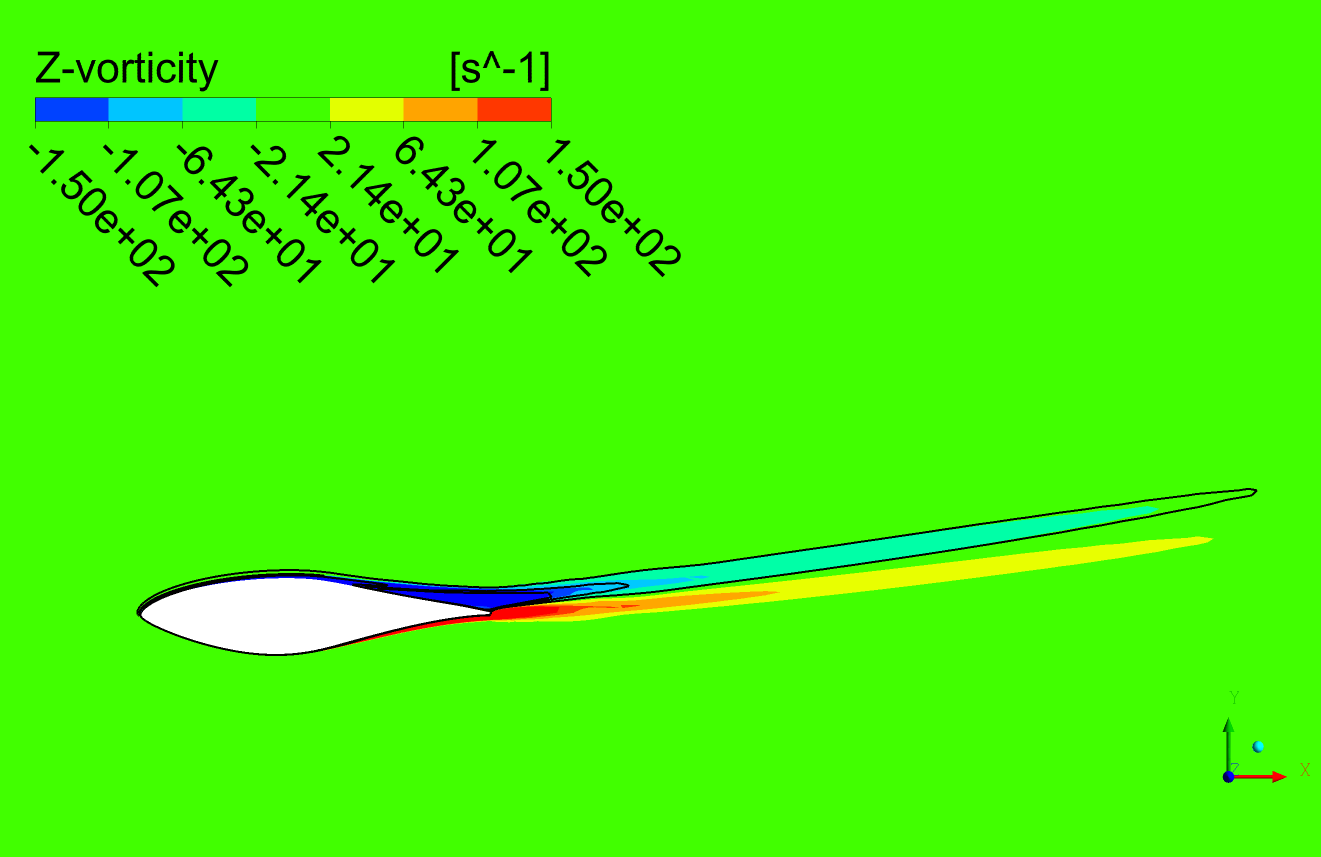}}\\
                \subfigure[$\alpha$=15$^{\circ}$; $c/r$=0.00; $Ro$=$\infty$ (without rotation)]{
                \includegraphics[width=0.34\textwidth]{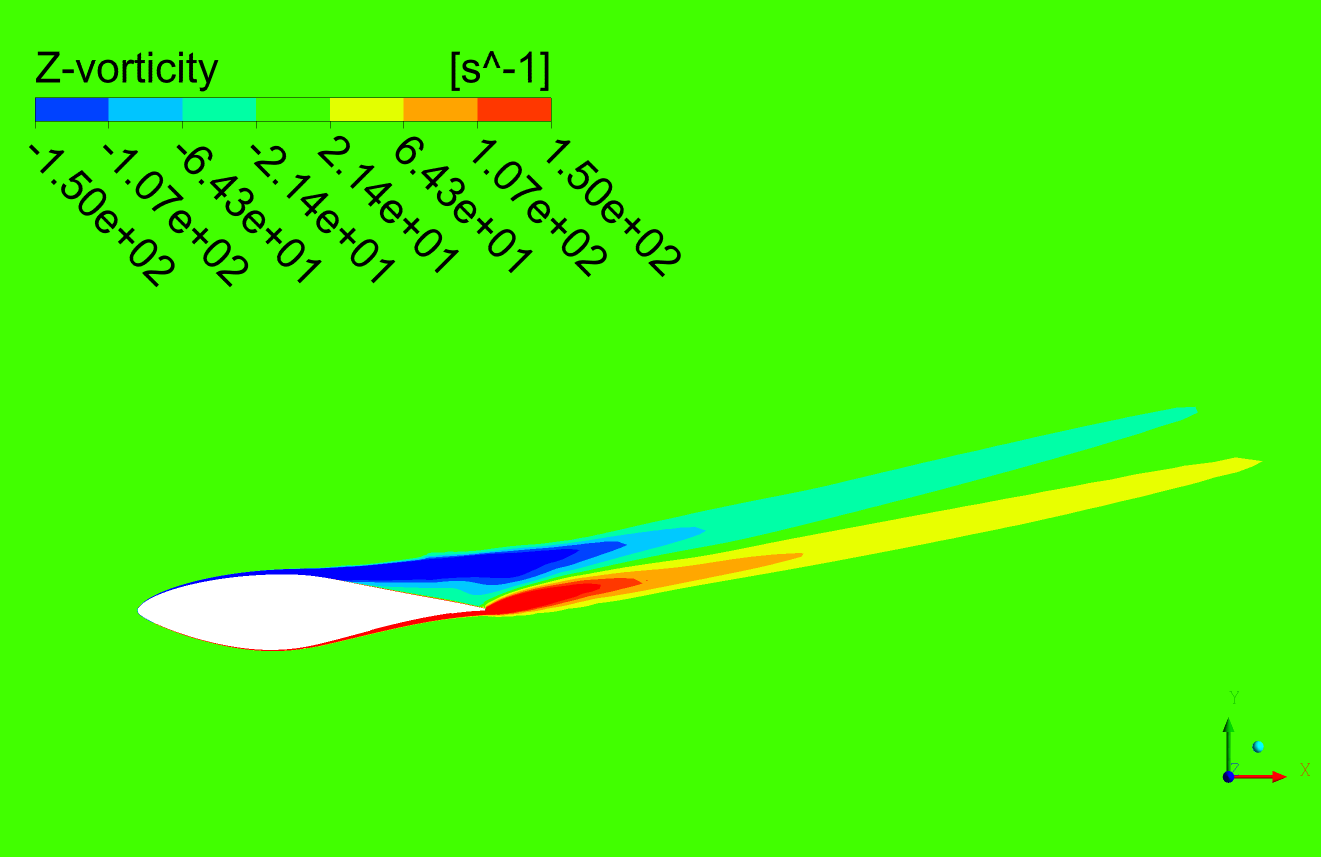}}
                \subfigure[$\alpha$=15$^{\circ}$; $c/r$=0.45; $Ro$=3 (with rotation)]{
                \includegraphics[width=0.34\textwidth]{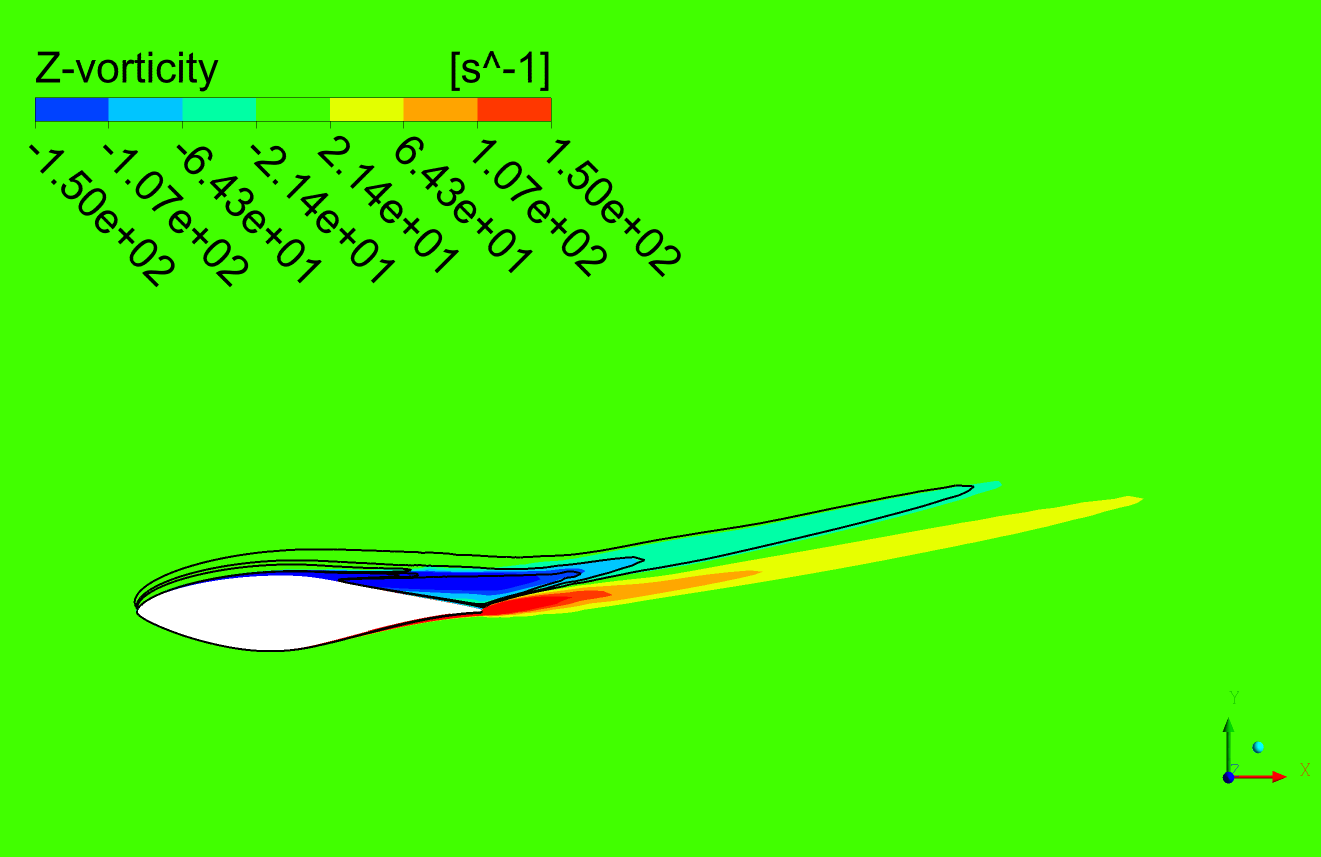}}\\              
                \subfigure[$\alpha$=27$^{\circ}$; $c/r$=0.0; $Ro$=$\infty$ (without rotation)]{
                 \includegraphics[width=0.31\textwidth]{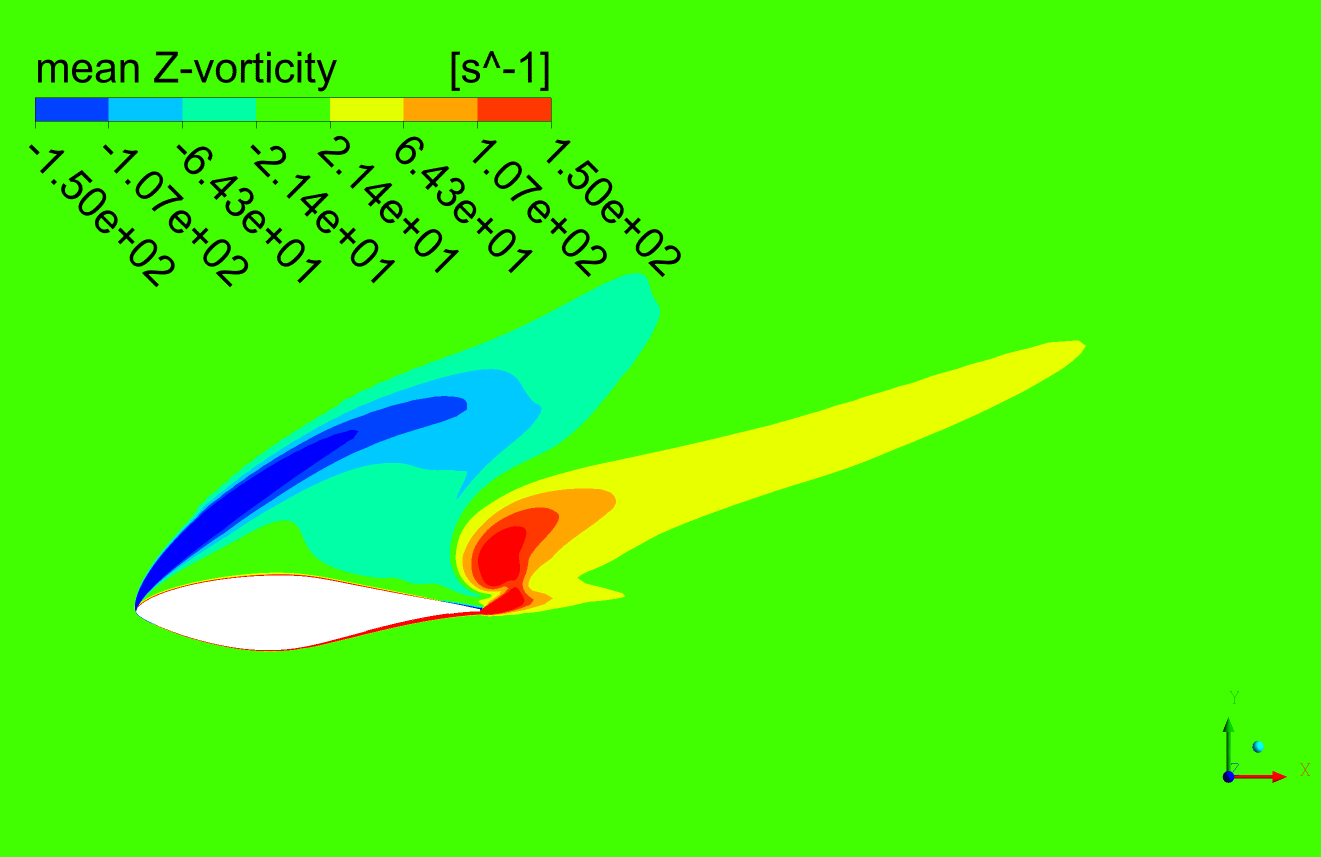}}
                \subfigure[$\alpha$=27$^{\circ}$; $c/r$=0.3; $Ro$=3.5 (with rotation)]{
                \includegraphics[width=0.31\textwidth]{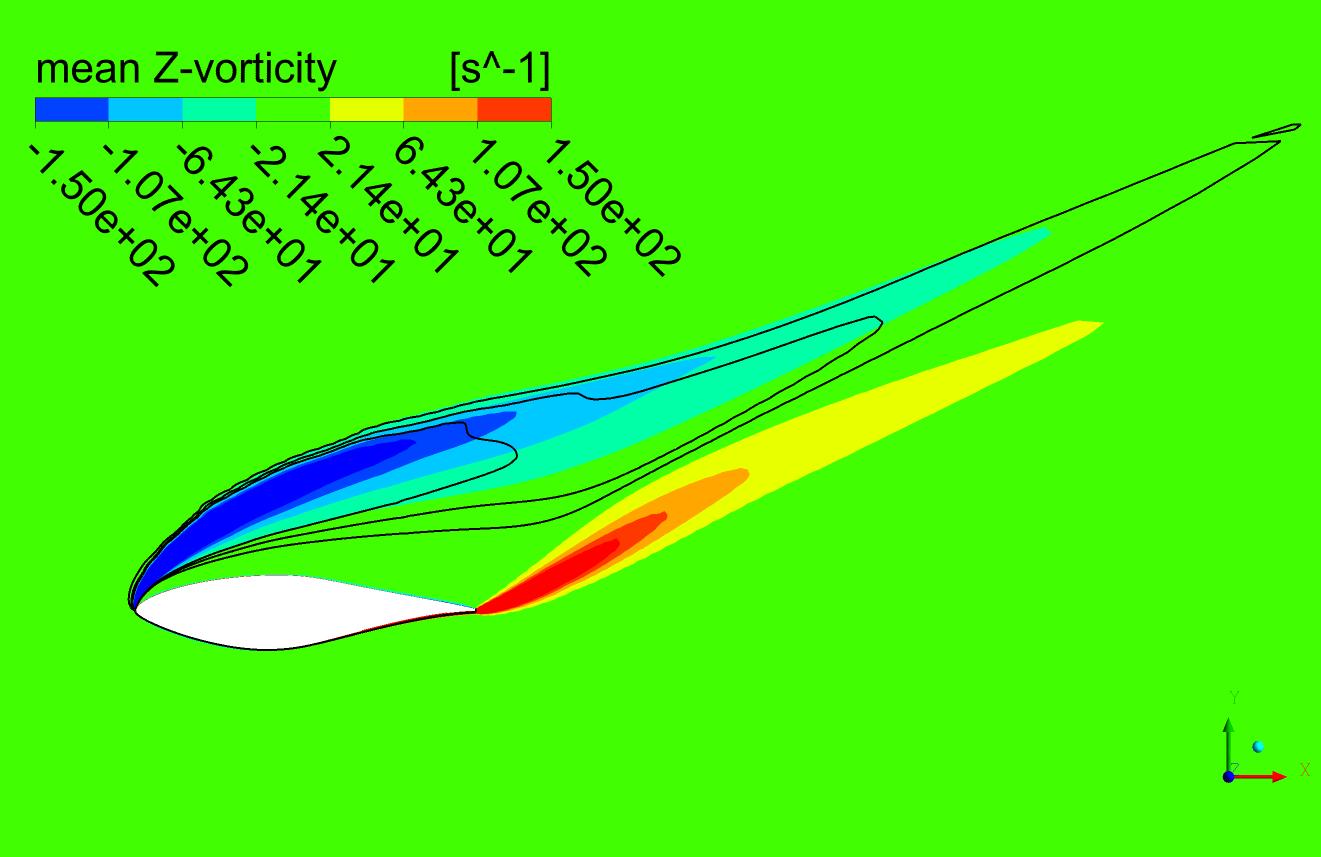}}
                \subfigure[$\alpha$=27$^{\circ}$; $c/r$=0.55; $Ro$=2 (with rotation)]{
                \includegraphics[width=0.31\textwidth]{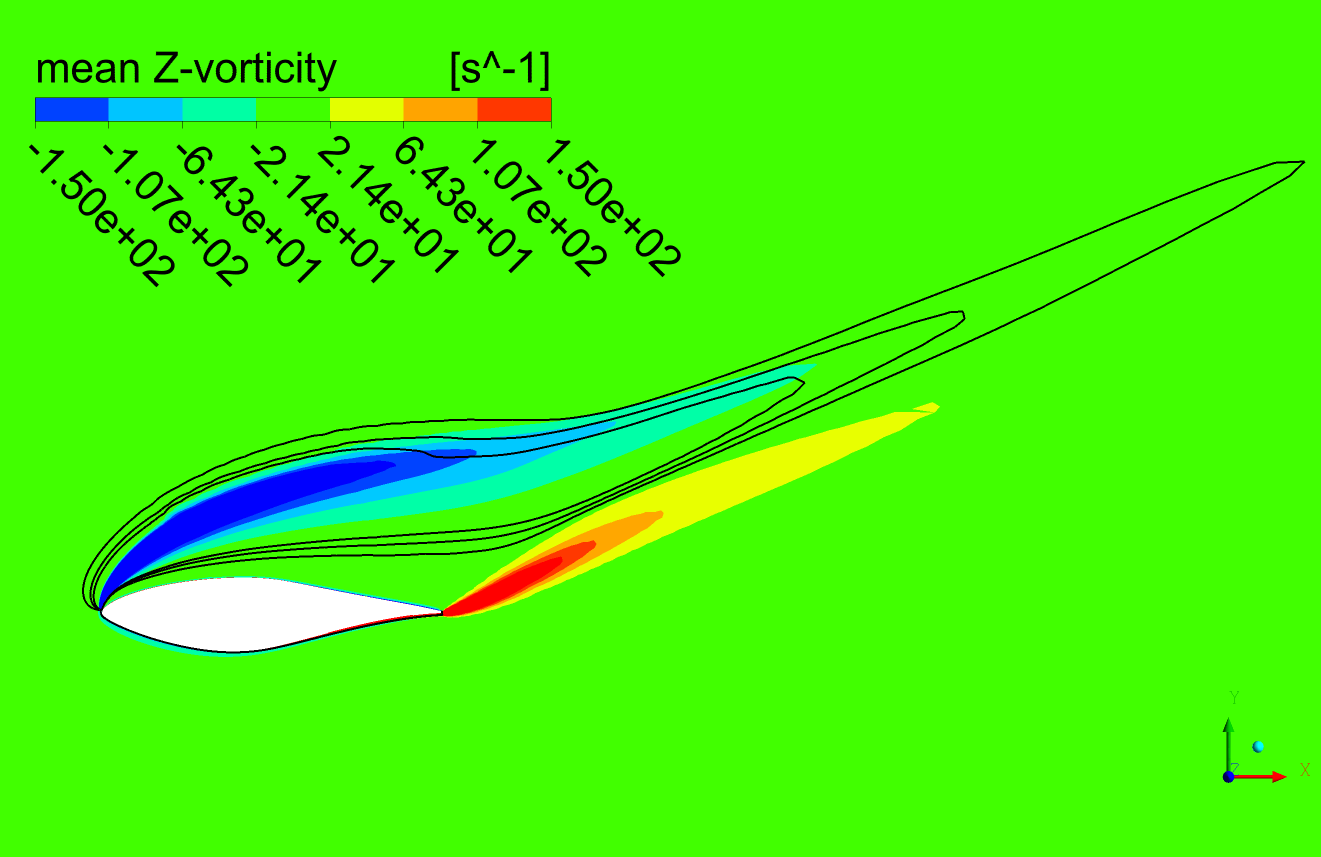}}
                \caption{Field of span-wise vorticity along with contours of the term associated to the planetary vorticity tilting, $S_{\omega}$, for the cases without (left-hand side) and with rotation (center and right-hand side) considering $\alpha$ = 10, 15 and 27 deg. The values of the $S_{\omega}$ contours are 7s$^{-2}$, 22s$^{-2}$ and 37s$^{-2}$ for $\alpha$ = 10 deg; 135s$^{-2}$, 285s$^{-2}$ and 435s$^{-2}$ for $\alpha$ = 15 deg; 185s$^{-2}$, 335s$^{-2}$ and 485s$^{-2}$ for $\alpha$ = 27 deg.}
                \label{fig:vortZ-cases}
            \end{figure*}

\noindent Note that $S_{\omega}$ is proportional to the gradient of $w$ in the direction of the axis of rotation. Figure \ref{fig:vortZ-cases} illustrates the radial vorticity together with contours of $S_{\omega}$ considering $\alpha$ = 10, 15 and 27 deg, respectively, for cases without and with rotation. It might be noted that high values of $S_{\omega}$ were found at the borders of the separate flow regions. As a result, $S_{\omega}$ counteracts the radial component of the vorticity coming in from the boundary layer and feeding the vortex, and thus tends to stabilize it \citep{werner:2019,eldredge:2019}.

\begin{figure*}
    \centering
    \subfigure[$\alpha$=10$^{\circ}$; $c/r$=0.0625; $Ro$=16]{
    \includegraphics[width=0.45\textwidth]{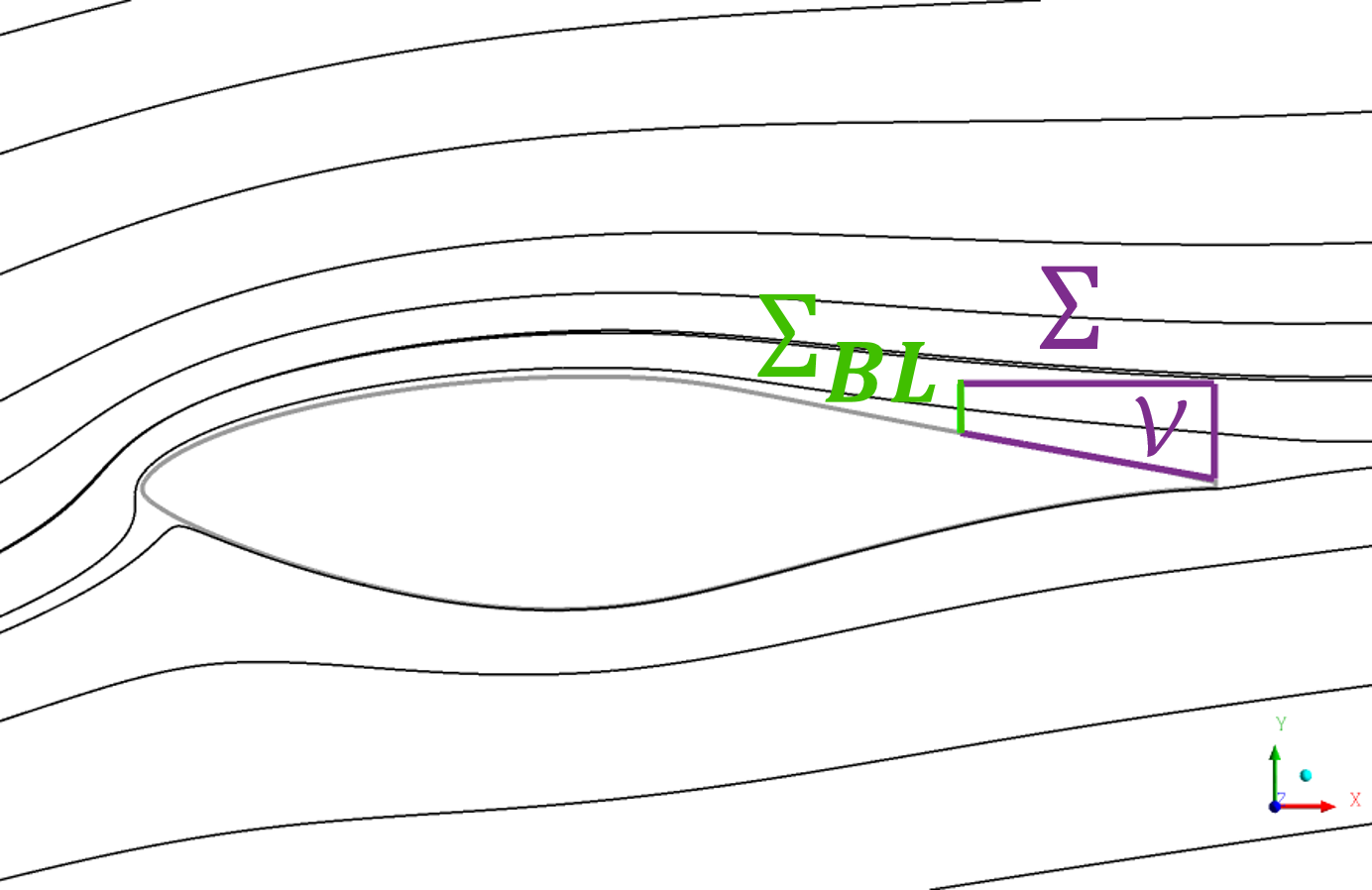}
    \label{fig:domain_alpha10}}
    \quad
    \subfigure[$\alpha$=15$^{\circ}$; $c/r$=0.45; $Ro$=3]{
    \includegraphics[width=0.45\textwidth]{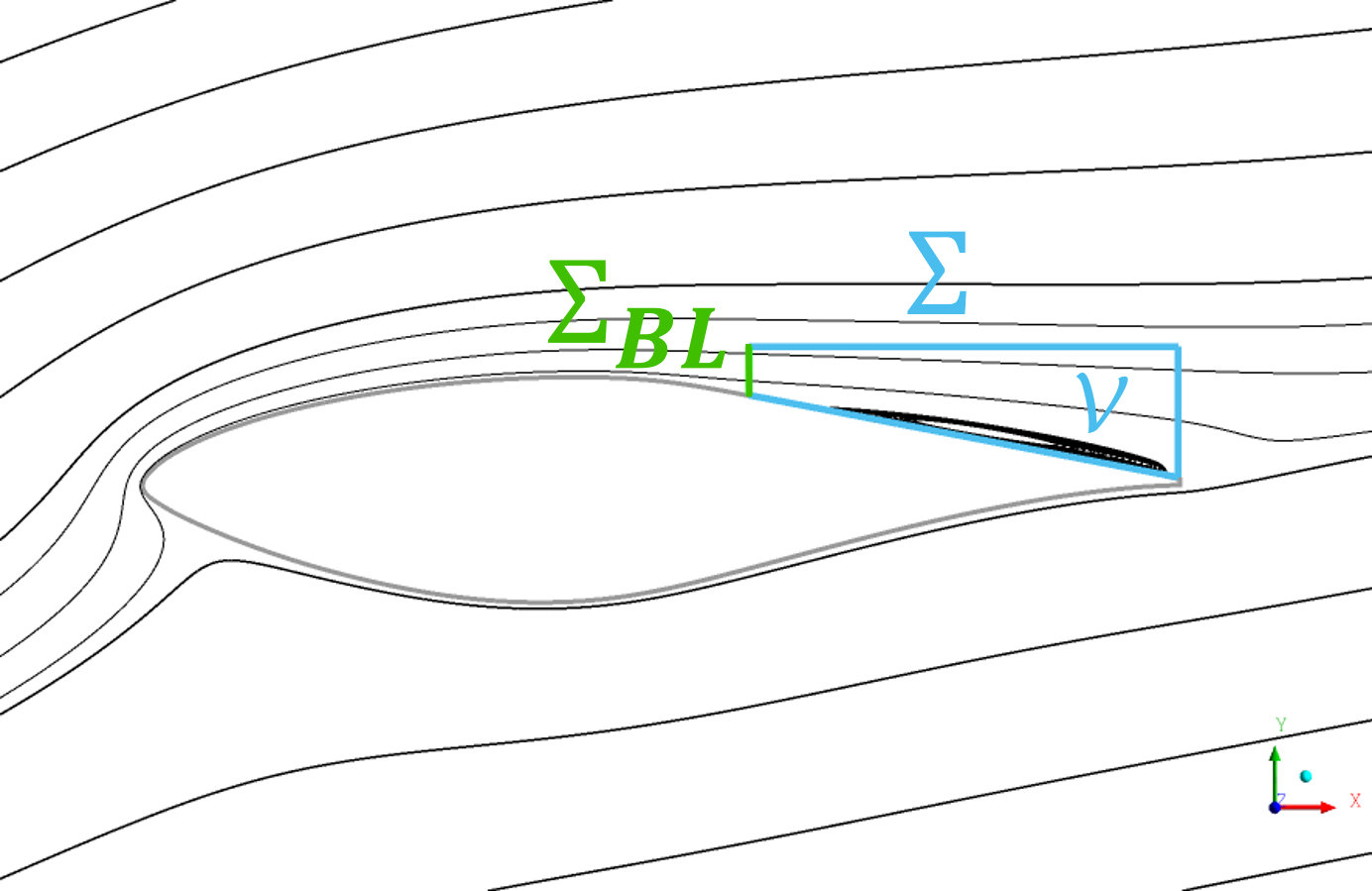}
    \label{fig:domain_alpha15}}
    \quad
    \subfigure[$\alpha$=27$^{\circ}$; $c/r$=0.3; $Ro$=3.5]{
    \includegraphics[width=0.45\textwidth]{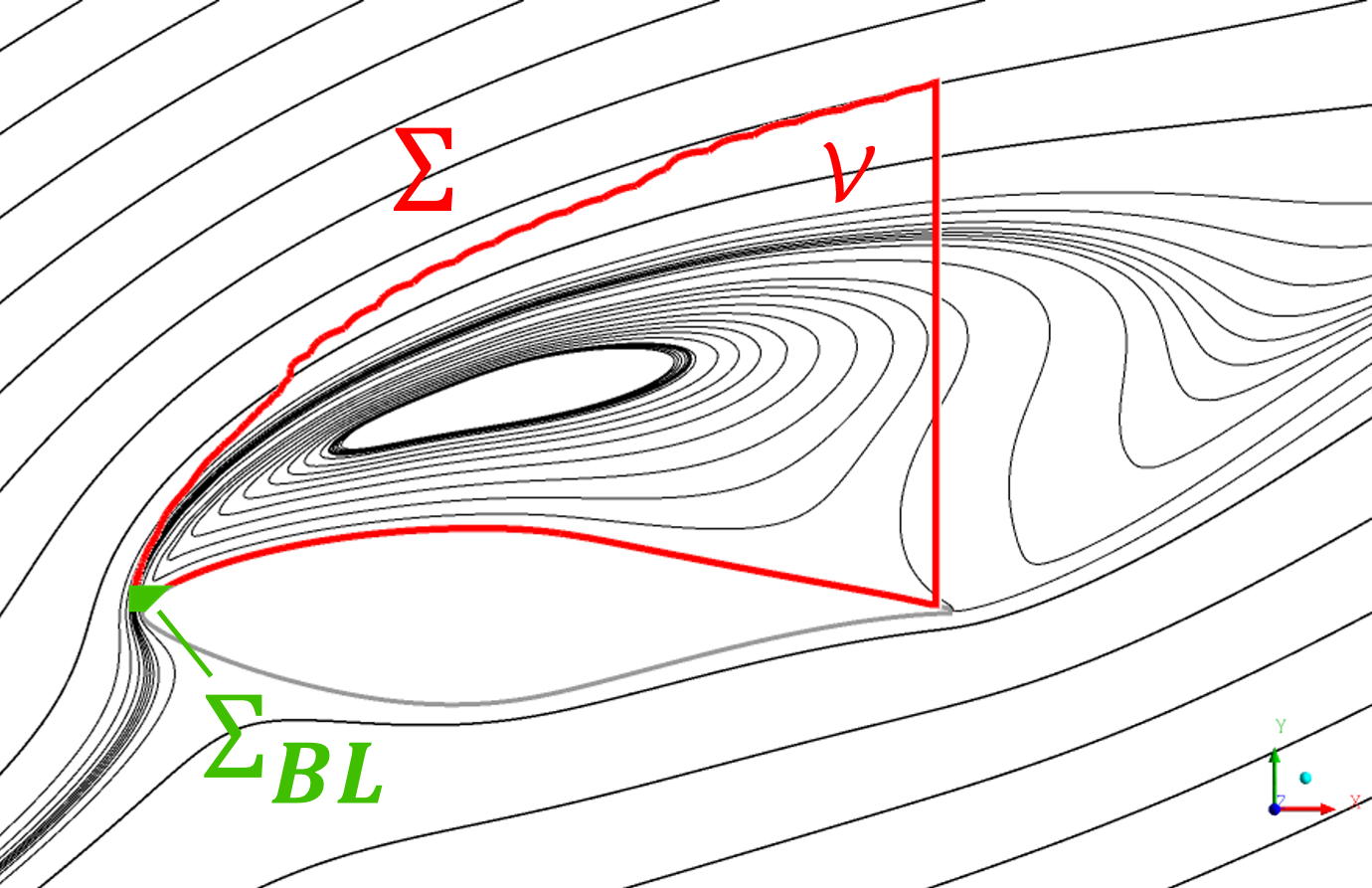}
    \label{fig:domain_alpha27_Ro3-5}}
    \quad
    \subfigure[$\alpha$=27$^{\circ}$; $c/r$=0.55; $Ro$=2]{
    \includegraphics[width=0.45\textwidth]{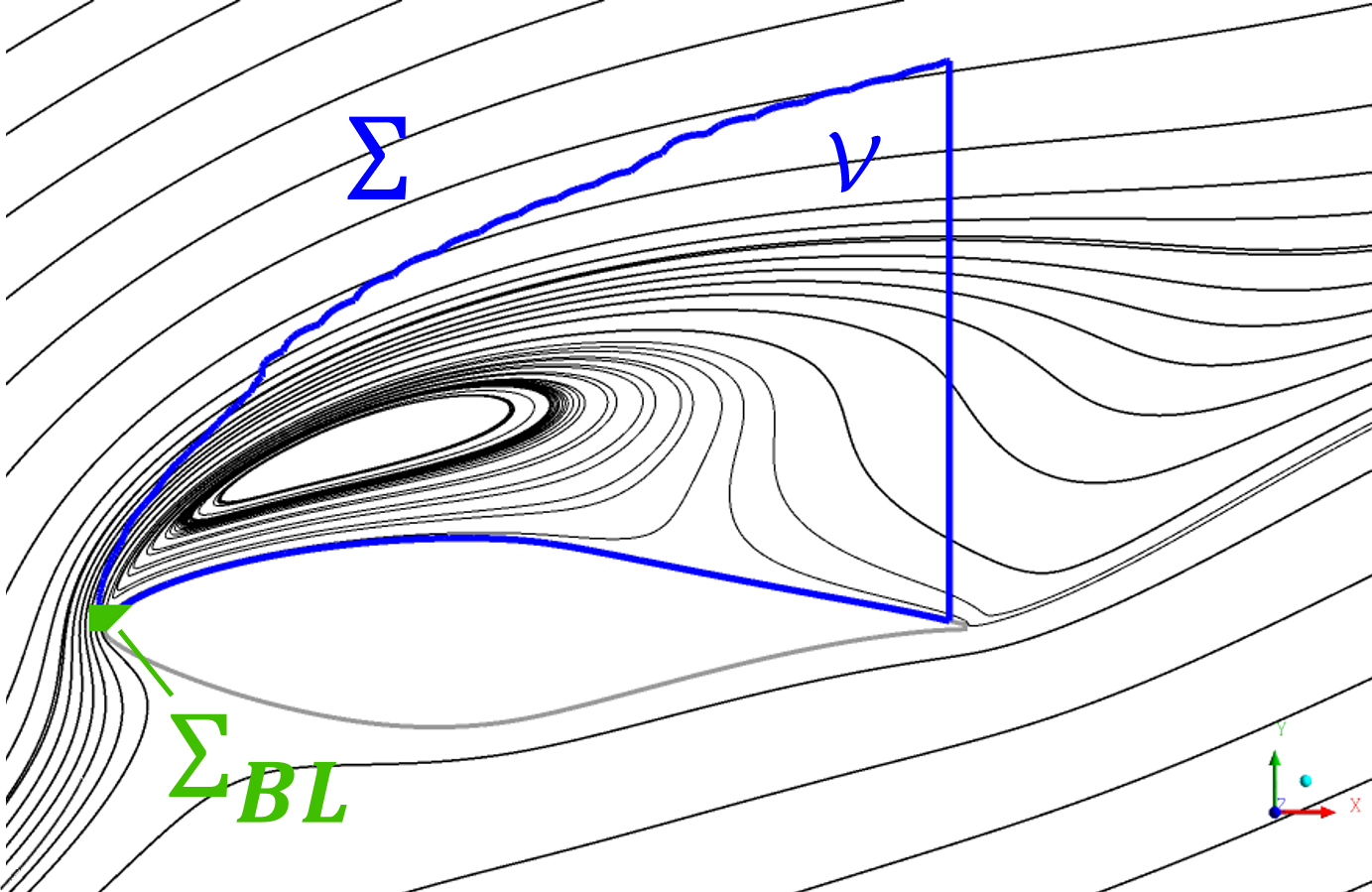}
    \label{fig:domain_alpha27_Ro02}}
    \caption{Integration domains of the Coriolis term of the radial vorticity equation. Limits based on the contours of the radial vorticity containing the recirculating region for each case.}
    \label{fig:domain_alpha}
\end{figure*}

To assess the relevance of $S_{\omega}$ to the stabilization of the separate flow, we compared its contribution to the balance of $\omega_z$, in a region containing the recirculation,

\begin{equation}
\overline{S}=\iint_{\mathcal{V}} S_{\omega} \ d\mathcal{V} = -2 \Omega \oint_{\Sigma} w \ \mathbf{e}_a \cdot \mathbf{n} \ d\Sigma, \nonumber
\end{equation}

\noindent with the contribution of the vorticity convecting from the boundary layer,

\begin{equation}
\dot{\omega}_{z,BL} = \int_{\Sigma_{BL}} \omega_z \mathbf{V} \cdot \mathbf{n} \ d\Sigma, \nonumber
\end{equation}

\noindent where $\mathbf{e}_a=(\sin\varphi, \cos\varphi)$ is the unit vector in the axial direction and $\mathbf{n}$ is the unit vector normal to the region boundary.
Four cases were considered in these analyses and the integration domains are shown in Figure \ref{fig:domain_alpha}.

Table \ref{tab:PVTr} presents the ratio obtained between the integral of $S_{\omega}$ and $\dot{\omega}_{z,BL}$ for cases with rotation.
It is observed that the relative contribution of $S_{\omega}$ correlated well with the effect of rotation on lift, i.e. it was negligible for the angle of attack of 10 deg, mild for the case with $\alpha$ = 15 deg and more intense for the cases with 27 deg angle of attack. For sufficiently high angle of attack, the radial planetary vorticity tilting may thus have a significant impact on the radial vorticity budget of the region of separate flow. In addition to other mechanisms, such as the centrifugal pumping, it may have a relevant contribution to reduce the volume of the region of separate flow or to stabilize the leading-edge vortex, and thus to a substantial lift augmentation.

\begin{table}[H]
    \centering
    \caption{Ratios between the integral of $S_\omega$ and rate of radial vorticity flowing in from the boundary layer.}
    \begin{ruledtabular}
    \begin{tabular}{clcc}
        $\alpha$ (deg) & \enspace $c/r$ & $Ro$ & \enspace $\overline{S}/\dot{\omega_{BL}}$ (\%) \\
        \hline
        10 & 0.0625 & 16 & 0.3495 \\ 
        15 & 0.45 & 3 & 8.464 \\ 
        27 & 0.3 & 3.5 & 40.32 \\
        27 & 0.55 & 2 & 41.93 \\
    \end{tabular}
    \end{ruledtabular}
    \label{tab:PVTr}
\end{table}

\newpage

\section{Conclusions}

Quasi-3D simulations were performed to explore the flow characteristics over sections of horizontal-axis wind turbines under rotation. The primary goal was to elucidate the influence of the tangential component of the Coriolis acceleration on the balance of radial vorticity at elevated angles of attack.
Comparisons between simulations without rotation and 2D wind-tunnel experiments demonstrated favorable correlations. Furthermore, analyses were conducted under conditions pertinent to the blades of horizontal-axis wind turbines.
For the cases with rotation, the simulations did not reproduce exactly experimental pressure distributions. However, they were capable of capturing important effects of rotation in qualitative correspondence with experiments and higher fidelity simulations, and provided interesting insights about the underlying physics of the rotational augmentation.

Within the conditions considered, the Rossby number had a greater impact than the $c/r$ ratio.
Three different regimes of rotational effect could be identified. At slightly high angles of attack, around 10 deg, the rotation had little influence on the aerodynamics and on the flow field. At mildly high angles of attack, around 15 deg, for which a trailing-edge separation was observed, the rotation reduced the volume of the region of separate flow, probably weakening the decambering effect, and thus increasing lift. This observation is in accordance with three-dimensional simulations of a complete HAWT blade. At a higher angle of attack, of 27 deg, for which the flow featured a leading-edge separation, for sufficiently low Rossby numbers, the rotation stabilized a leading-edge vortex and caused a substantial lift increase.

The rate of destruction of span-wise vorticity by the Coriolis force in the region of separate flow was compared to the vorticity inflow from the upstream boundary layer. Particularly, for conditions at 27 deg angle of attack, the Coriolis force arguably destroyed around 40\% of the span-wise vorticity flowing in from the boundary layer, which suggests that it has a significant contribution to the stabilization of the leading-edge vortex.
Since, for small wind turbines, large rotational augmentation was associated to the presence of a stationary vortex on the upper surface, this so-called radial planetary vorticity tilting may be of great relevance for the aerodynamics of HAWT blades.

\newpage

\begin{acknowledgments}
This research was supported by FAPESP under grants \#2020/10972-3 and \#2022/02812-1, São Paulo Research Foundation (FAPESP).
\end{acknowledgments}

\newpage

\section*{Data Availability Statement}

The data that support the findings of this study are available from the corresponding author upon reasonable request.

\newpage

\section*{Author Declarations}

\subsection*{Conflict of Interest}
The authors have no conflicts to disclose.

\subsection*{Author Contributions}
\noindent \textbf{Pedro Rodrigues:} Conceptualization (supporting); Data curation (lead); Formal analysis (equal); Funding acquisition (supporting); Investigation (lead); Methodology (lead); Resources (supporting); Software (lead); Validation (lead); Visualization (equal); Writing – original draft (lead); Writing – review \& editing (equal). \textbf{Elmer Gennaro:} Conceptualization (supporting); Validation (supporting); Writing – review \& editing (equal). \textbf{Daniel Souza:} Conceptualization (lead); Data curation (supporting); Formal analysis (equal); Funding acquisition (lead); Investigation (supporting); Methodology (supporting); Project administration (lead); Resources (lead); Software (supporting); Supervision (lead); Validation (supporting); Visualization (equal); Writing – original draft (supporting); Writing – review \& editing (equal).

\section*{References}

\bibliography{references}

%merlin.mbs aipnum4-1.bst 2010-07-25 4.21a (PWD, AO, DPC) hacked
%Control: key (0)
%Control: author (8) initials jnrlst
%Control: editor formatted (1) identically to author
%Control: production of article title (0) allowed
%Control: page (1) range
%Control: year (1) truncated
%Control: production of eprint (0) enabled
\begin{thebibliography}{25}%
\makeatletter
\providecommand \@ifxundefined [1]{%
 \@ifx{#1\undefined}
}%
\providecommand \@ifnum [1]{%
 \ifnum #1\expandafter \@firstoftwo
 \else \expandafter \@secondoftwo
 \fi
}%
\providecommand \@ifx [1]{%
 \ifx #1\expandafter \@firstoftwo
 \else \expandafter \@secondoftwo
 \fi
}%
\providecommand \natexlab [1]{#1}%
\providecommand \enquote  [1]{``#1''}%
\providecommand \bibnamefont  [1]{#1}%
\providecommand \bibfnamefont [1]{#1}%
\providecommand \citenamefont [1]{#1}%
\providecommand \href@noop [0]{\@secondoftwo}%
\providecommand \href [0]{\begingroup \@sanitize@url \@href}%
\providecommand \@href[1]{\@@startlink{#1}\@@href}%
\providecommand \@@href[1]{\endgroup#1\@@endlink}%
\providecommand \@sanitize@url [0]{\catcode `\\12\catcode `\$12\catcode
  `\&12\catcode `\#12\catcode `\^12\catcode `\_12\catcode `\%12\relax}%
\providecommand \@@startlink[1]{}%
\providecommand \@@endlink[0]{}%
\providecommand \url  [0]{\begingroup\@sanitize@url \@url }%
\providecommand \@url [1]{\endgroup\@href {#1}{\urlprefix }}%
\providecommand \urlprefix  [0]{URL }%
\providecommand \Eprint [0]{\href }%
\providecommand \doibase [0]{http://dx.doi.org/}%
\providecommand \selectlanguage [0]{\@gobble}%
\providecommand \bibinfo  [0]{\@secondoftwo}%
\providecommand \bibfield  [0]{\@secondoftwo}%
\providecommand \translation [1]{[#1]}%
\providecommand \BibitemOpen [0]{}%
\providecommand \bibitemStop [0]{}%
\providecommand \bibitemNoStop [0]{.\EOS\space}%
\providecommand \EOS [0]{\spacefactor3000\relax}%
\providecommand \BibitemShut  [1]{\csname bibitem#1\endcsname}%
\let\auto@bib@innerbib\@empty
%</preamble>
\bibitem [{\citenamefont {Dwyer}\ and\ \citenamefont
  {McCroskey}(1970)}]{dwyer:1970}%
  \BibitemOpen
  \bibfield  {author} {\bibinfo {author} {\bibfnamefont {H.}~\bibnamefont
  {Dwyer}}\ and\ \bibinfo {author} {\bibfnamefont {W.}~\bibnamefont
  {McCroskey}},\ }\bibfield  {title} {\enquote {\bibinfo {title} {Crossflow and
  unsteady boundary-layer effects on rotating blades},}\ }in\ \href@noop {}
  {\emph {\bibinfo {booktitle} {Proceedings of 8th AIAA Aerospace Science
  Meeting}}}\ (\bibinfo {address} {New York, USA},\ \bibinfo {year}
  {1970})\BibitemShut {NoStop}%
\bibitem [{\citenamefont {Ronsten}(1992)}]{ronsten:1992}%
  \BibitemOpen
  \bibfield  {author} {\bibinfo {author} {\bibfnamefont {G.}~\bibnamefont
  {Ronsten}},\ }\bibfield  {title} {\enquote {\bibinfo {title} {Static pressure
  measurements on a rotating and a non-rotating 2.375 m wind turbine blade.
  {C}omparison with {2D} calculations},}\ }\href@noop {} {\bibfield  {journal}
  {\bibinfo  {journal} {Journal of Wind Engineering and Industrial
  Aerodynamics}\ }\textbf {\bibinfo {volume} {39}},\ \bibinfo {pages}
  {105--118} (\bibinfo {year} {1992})}\BibitemShut {NoStop}%
\bibitem [{\citenamefont {Du}\ and\ \citenamefont {Selig}(2000)}]{du:2000}%
  \BibitemOpen
  \bibfield  {author} {\bibinfo {author} {\bibfnamefont {Z.}~\bibnamefont
  {Du}}\ and\ \bibinfo {author} {\bibfnamefont {M.}~\bibnamefont {Selig}},\
  }\bibfield  {title} {\enquote {\bibinfo {title} {The effect of rotation on
  the boundary layer of a wind turbine},}\ }\href@noop {} {\bibfield  {journal}
  {\bibinfo  {journal} {Renewable Energy}\ }\textbf {\bibinfo {volume} {20}},\
  \bibinfo {pages} {167--181} (\bibinfo {year} {2000})}\BibitemShut {NoStop}%
\bibitem [{\citenamefont {Mauro}, \citenamefont {Lanzafame},\ and\
  \citenamefont {Messina}(2017{\natexlab{a}})}]{mauro2:2017}%
  \BibitemOpen
  \bibfield  {author} {\bibinfo {author} {\bibfnamefont {S.}~\bibnamefont
  {Mauro}}, \bibinfo {author} {\bibfnamefont {R.}~\bibnamefont {Lanzafame}}, \
  and\ \bibinfo {author} {\bibfnamefont {M.}~\bibnamefont {Messina}},\
  }\bibfield  {title} {\enquote {\bibinfo {title} {An insight into the
  rotational augmentation on {HAWT}s by means of {CFD} simulations - {PART II}:
  Post processing and force analysis},}\ }\href@noop {} {\bibfield  {journal}
  {\bibinfo  {journal} {International Journal of Applied Engineering Research}\
  }\textbf {\bibinfo {volume} {12}},\ \bibinfo {pages} {10505--10529} (\bibinfo
  {year} {2017}{\natexlab{a}})}\BibitemShut {NoStop}%
\bibitem [{\citenamefont {Sicot}\ \emph {et~al.}(2008)\citenamefont {Sicot},
  \citenamefont {Devinant}, \citenamefont {Loyer},\ and\ \citenamefont
  {Hureau}}]{sicot:2008}%
  \BibitemOpen
  \bibfield  {author} {\bibinfo {author} {\bibfnamefont {C.}~\bibnamefont
  {Sicot}}, \bibinfo {author} {\bibfnamefont {P.}~\bibnamefont {Devinant}},
  \bibinfo {author} {\bibfnamefont {S.}~\bibnamefont {Loyer}}, \ and\ \bibinfo
  {author} {\bibfnamefont {J.}~\bibnamefont {Hureau}},\ }\bibfield  {title}
  {\enquote {\bibinfo {title} {Rotational and turbulence effect on a wind
  turbine blade. {I}nvestigation of the stall mechanisms},}\ }\href@noop {}
  {\bibfield  {journal} {\bibinfo  {journal} {Journal of Wind Engineering and
  Industrial Aerodynamics}\ }\textbf {\bibinfo {volume} {96}},\ \bibinfo
  {pages} {1320--1331} (\bibinfo {year} {2008})}\BibitemShut {NoStop}%
\bibitem [{\citenamefont {Schreck}, \citenamefont {Soerensen},\ and\
  \citenamefont {Robinson}(2007)}]{schreck:2007}%
  \BibitemOpen
  \bibfield  {author} {\bibinfo {author} {\bibfnamefont {S.}~\bibnamefont
  {Schreck}}, \bibinfo {author} {\bibfnamefont {N.}~\bibnamefont {Soerensen}},
  \ and\ \bibinfo {author} {\bibfnamefont {M.}~\bibnamefont {Robinson}},\
  }\bibfield  {title} {\enquote {\bibinfo {title} {Aerodynamic structures and
  processes in rotationally augmented flow fields},}\ }\href@noop {} {\bibfield
   {journal} {\bibinfo  {journal} {Wind Energy}\ }\textbf {\bibinfo {volume}
  {10}},\ \bibinfo {pages} {159--178} (\bibinfo {year} {2007})}\BibitemShut
  {NoStop}%
\bibitem [{\citenamefont {Jardin}\ and\ \citenamefont
  {David}(2014)}]{jardin:2014}%
  \BibitemOpen
  \bibfield  {author} {\bibinfo {author} {\bibfnamefont {T.}~\bibnamefont
  {Jardin}}\ and\ \bibinfo {author} {\bibfnamefont {L.}~\bibnamefont {David}},\
  }\bibfield  {title} {\enquote {\bibinfo {title} {Spanwise gradients in flow
  speed help stabilize leading-edge vortices on revolving wings},}\ }\href@noop
  {} {\bibfield  {journal} {\bibinfo  {journal} {Physical Review E}\ }\textbf
  {\bibinfo {volume} {90}},\ \bibinfo {pages} {312--340} (\bibinfo {year}
  {2014})}\BibitemShut {NoStop}%
\bibitem [{\citenamefont {Corten}(2001)}]{corten:2001}%
  \BibitemOpen
  \bibfield  {author} {\bibinfo {author} {\bibfnamefont {G.}~\bibnamefont
  {Corten}},\ }\bibfield  {title} {\enquote {\bibinfo {title} {Flow separation
  on wind turbine blades},}\ \ }(\bibinfo {address} {Ultrecht, Netherlands},\
  \bibinfo {year} {2001})\BibitemShut {NoStop}%
\bibitem [{\citenamefont {Lindenburg}(2003)}]{lindenburg:2003}%
  \BibitemOpen
  \bibfield  {author} {\bibinfo {author} {\bibfnamefont {C.}~\bibnamefont
  {Lindenburg}},\ }\href@noop {} {\enquote {\bibinfo {title} {Investigation
  into rotor blade aerodynamics},}\ }\bibinfo {type} {Tech. Rep.}\ (\bibinfo
  {institution} {NOVEM},\ \bibinfo {address} {Netherlands},\ \bibinfo {year}
  {2003})\BibitemShut {NoStop}%
\bibitem [{\citenamefont {Bangga}\ \emph {et~al.}(2017)\citenamefont {Bangga},
  \citenamefont {Lutz}, \citenamefont {Jost},\ and\ \citenamefont
  {Kr\"amer}}]{bangga:2017b}%
  \BibitemOpen
  \bibfield  {author} {\bibinfo {author} {\bibfnamefont {G.}~\bibnamefont
  {Bangga}}, \bibinfo {author} {\bibfnamefont {T.}~\bibnamefont {Lutz}},
  \bibinfo {author} {\bibfnamefont {E.}~\bibnamefont {Jost}}, \ and\ \bibinfo
  {author} {\bibfnamefont {E.}~\bibnamefont {Kr\"amer}},\ }\bibfield  {title}
  {\enquote {\bibinfo {title} {{CFD} studies on rotational augmentation at the
  inboard sections of a 10{MW} wind turbine rotor},}\ }\href@noop {} {\bibfield
   {journal} {\bibinfo  {journal} {Journal of Renewable and Sustainable
  Energy}\ }\textbf {\bibinfo {volume} {9}},\ \bibinfo {pages} {1--28}
  (\bibinfo {year} {2017})}\BibitemShut {NoStop}%
\bibitem [{\citenamefont {Dumitrescu}, \citenamefont {Cardos},\ and\
  \citenamefont {Dumitrache}(2007)}]{dumitrescu:2007}%
  \BibitemOpen
  \bibfield  {author} {\bibinfo {author} {\bibfnamefont {H.}~\bibnamefont
  {Dumitrescu}}, \bibinfo {author} {\bibfnamefont {V.}~\bibnamefont {Cardos}},
  \ and\ \bibinfo {author} {\bibfnamefont {A.}~\bibnamefont {Dumitrache}},\
  }\bibfield  {title} {\enquote {\bibinfo {title} {Modelling of inboard stall
  delay due to rotaion},}\ }\href@noop {} {\bibfield  {journal} {\bibinfo
  {journal} {Journal of Physics: Conference Series}\ }\textbf {\bibinfo
  {volume} {75}} (\bibinfo {year} {2007})}\BibitemShut {NoStop}%
\bibitem [{\citenamefont {Polhamus}(1966)}]{polhamus:1966}%
  \BibitemOpen
  \bibfield  {author} {\bibinfo {author} {\bibfnamefont {E.}~\bibnamefont
  {Polhamus}},\ }\href@noop {} {\enquote {\bibinfo {title} {A concept of vortex
  lift of sharp-edge wings based on a leading-edge-suction analogy},}\
  }\bibinfo {type} {Tech. Rep.}\ (\bibinfo  {institution} {NASA},\ \bibinfo
  {address} {Hampton, Virginia},\ \bibinfo {year} {1966})\BibitemShut {NoStop}%
\bibitem [{\citenamefont {Wojcik}\ and\ \citenamefont
  {Buchholtz}(2014)}]{wojcik:2014}%
  \BibitemOpen
  \bibfield  {author} {\bibinfo {author} {\bibfnamefont {C.}~\bibnamefont
  {Wojcik}}\ and\ \bibinfo {author} {\bibfnamefont {J.}~\bibnamefont
  {Buchholtz}},\ }\bibfield  {title} {\enquote {\bibinfo {title} {Vorticity
  transport in the leading-edge vortex on a rotating blade},}\ }\href@noop {}
  {\bibfield  {journal} {\bibinfo  {journal} {Journal of Fluid Mechanics}\
  }\textbf {\bibinfo {volume} {743}},\ \bibinfo {pages} {249--261} (\bibinfo
  {year} {2014})}\BibitemShut {NoStop}%
\bibitem [{\citenamefont {Jardin}(2017)}]{jardin:2017}%
  \BibitemOpen
  \bibfield  {author} {\bibinfo {author} {\bibfnamefont {T.}~\bibnamefont
  {Jardin}},\ }\bibfield  {title} {\enquote {\bibinfo {title} {Coriolis effect
  and the attachment of the leading edge vortex},}\ }\href@noop {} {\bibfield
  {journal} {\bibinfo  {journal} {Journal of Fluid Mechanics}\ }\textbf
  {\bibinfo {volume} {820}},\ \bibinfo {pages} {312--340} (\bibinfo {year}
  {2017})}\BibitemShut {NoStop}%
\bibitem [{\citenamefont {Werner}\ \emph {et~al.}(2019)\citenamefont {Werner},
  \citenamefont {Chung}, \citenamefont {Wang}, \citenamefont {Liu},
  \citenamefont {Cimbala}, \citenamefont {Dong},\ and\ \citenamefont
  {Cheng}}]{werner:2019}%
  \BibitemOpen
  \bibfield  {author} {\bibinfo {author} {\bibfnamefont {N.}~\bibnamefont
  {Werner}}, \bibinfo {author} {\bibfnamefont {H.}~\bibnamefont {Chung}},
  \bibinfo {author} {\bibfnamefont {J.}~\bibnamefont {Wang}}, \bibinfo {author}
  {\bibfnamefont {G.}~\bibnamefont {Liu}}, \bibinfo {author} {\bibfnamefont
  {J.}~\bibnamefont {Cimbala}}, \bibinfo {author} {\bibfnamefont
  {H.}~\bibnamefont {Dong}}, \ and\ \bibinfo {author} {\bibfnamefont
  {B.}~\bibnamefont {Cheng}},\ }\bibfield  {title} {\enquote {\bibinfo {title}
  {Radial planetary vorticity tilting in the leading-edge vortex of revolving
  wings},}\ }\href@noop {} {\bibfield  {journal} {\bibinfo  {journal} {Physics
  of Fluids}\ }\textbf {\bibinfo {volume} {31}},\ \bibinfo {pages} {1--15}
  (\bibinfo {year} {2019})}\BibitemShut {NoStop}%
\bibitem [{\citenamefont {Souza}\ and\ \citenamefont
  {Gennaro}(2020)}]{souza:2020}%
  \BibitemOpen
  \bibfield  {author} {\bibinfo {author} {\bibfnamefont {D.}~\bibnamefont
  {Souza}}\ and\ \bibinfo {author} {\bibfnamefont {E.}~\bibnamefont
  {Gennaro}},\ }\bibfield  {title} {\enquote {\bibinfo {title} {Rotational
  effects on the spanwise periodic flow over a wind turbine airfoil},}\
  }\href@noop {} {\bibfield  {journal} {\bibinfo  {journal} {18th Brazilian
  Congress of Thermal Sciences and Engineering}\ ,\ \bibinfo {pages} {1--11}}
  (\bibinfo {year} {2020})}\BibitemShut {NoStop}%
\bibitem [{\citenamefont {Chaviaropoulos}\ and\ \citenamefont
  {Hansen}(2000)}]{chaviaropoulos:2000}%
  \BibitemOpen
  \bibfield  {author} {\bibinfo {author} {\bibfnamefont {P.}~\bibnamefont
  {Chaviaropoulos}}\ and\ \bibinfo {author} {\bibfnamefont {M.}~\bibnamefont
  {Hansen}},\ }\bibfield  {title} {\enquote {\bibinfo {title} {Investigating
  three-dimensional and rotational effects on wind turbine blades by means of
  quasi-{3D} {Navier-Stokes} solver},}\ }\href@noop {} {\bibfield  {journal}
  {\bibinfo  {journal} {Journal of Fluids Engineering}\ }\textbf {\bibinfo
  {volume} {122}},\ \bibinfo {pages} {330--336} (\bibinfo {year}
  {2000})}\BibitemShut {NoStop}%
\bibitem [{\citenamefont {Patankar}(1980)}]{patankar:1980}%
  \BibitemOpen
  \bibfield  {author} {\bibinfo {author} {\bibfnamefont {S.~V.}\ \bibnamefont
  {Patankar}},\ }\href@noop {} {\emph {\bibinfo {title} {Numerical heat
  transfer and fluid flow}}}\ (\bibinfo  {publisher} {McGraw-Hill},\ \bibinfo
  {address} {Washington},\ \bibinfo {year} {1980})\ p.\ \bibinfo {pages}
  {193}\BibitemShut {NoStop}%
\bibitem [{\citenamefont {Mauro}, \citenamefont {Lanzafame},\ and\
  \citenamefont {Messina}(2017{\natexlab{b}})}]{mauro1:2017}%
  \BibitemOpen
  \bibfield  {author} {\bibinfo {author} {\bibfnamefont {S.}~\bibnamefont
  {Mauro}}, \bibinfo {author} {\bibfnamefont {R.}~\bibnamefont {Lanzafame}}, \
  and\ \bibinfo {author} {\bibfnamefont {M.}~\bibnamefont {Messina}},\
  }\bibfield  {title} {\enquote {\bibinfo {title} {An insight into the
  rotational augmentation on {HAWT}s by means of {CFD} simulations - {PART I}:
  State of the art and numerical results},}\ }\href@noop {} {\bibfield
  {journal} {\bibinfo  {journal} {International Journal of Applied Engineering
  Research}\ }\textbf {\bibinfo {volume} {12}},\ \bibinfo {pages}
  {10491--10504} (\bibinfo {year} {2017}{\natexlab{b}})}\BibitemShut {NoStop}%
\bibitem [{\citenamefont {Wang}\ \emph {et~al.}(2010)\citenamefont {Wang},
  \citenamefont {Ma}, \citenamefont {Ingham}, \citenamefont {Pourkashanian},\
  and\ \citenamefont {Tao}}]{wang:2010}%
  \BibitemOpen
  \bibfield  {author} {\bibinfo {author} {\bibfnamefont {S.}~\bibnamefont
  {Wang}}, \bibinfo {author} {\bibfnamefont {L.}~\bibnamefont {Ma}}, \bibinfo
  {author} {\bibfnamefont {D.}~\bibnamefont {Ingham}}, \bibinfo {author}
  {\bibfnamefont {M.}~\bibnamefont {Pourkashanian}}, \ and\ \bibinfo {author}
  {\bibfnamefont {Z.}~\bibnamefont {Tao}},\ }\bibfield  {title} {\enquote
  {\bibinfo {title} {Turbulence modeling of deep dynamic stall at low
  {Reynolds} number},}\ }in\ \href@noop {} {\emph {\bibinfo {booktitle}
  {Proceedings of the World Congress on Engineering 2010 Vol {II}}}}\ (\bibinfo
  {address} {Londres, Reino Unido},\ \bibinfo {year} {2010})\BibitemShut
  {NoStop}%
\bibitem [{\citenamefont {Menter}(1994)}]{menter:1994}%
  \BibitemOpen
  \bibfield  {author} {\bibinfo {author} {\bibfnamefont {F.~R.}\ \bibnamefont
  {Menter}},\ }\bibfield  {title} {\enquote {\bibinfo {title} {Two-equation
  eddy-viscosity turbulence models for engineering applications},}\ }\href@noop
  {} {\bibfield  {journal} {\bibinfo  {journal} {AIAA Journal}\ }\textbf
  {\bibinfo {volume} {32}},\ \bibinfo {pages} {1598--1605} (\bibinfo {year}
  {1994})}\BibitemShut {NoStop}%
\bibitem [{\citenamefont {Hand}\ \emph {et~al.}(2001)\citenamefont {Hand},
  \citenamefont {D.}, \citenamefont {Fingersh}, \citenamefont {Jager},
  \citenamefont {Cotrell},\ and\ \citenamefont {Larwood}}]{hand:2001}%
  \BibitemOpen
  \bibfield  {author} {\bibinfo {author} {\bibfnamefont {M.}~\bibnamefont
  {Hand}}, \bibinfo {author} {\bibfnamefont {S.}~\bibnamefont {D.}}, \bibinfo
  {author} {\bibfnamefont {L.}~\bibnamefont {Fingersh}}, \bibinfo {author}
  {\bibfnamefont {D.}~\bibnamefont {Jager}}, \bibinfo {author} {\bibfnamefont
  {S.}~\bibnamefont {Cotrell}, \bibfnamefont {J.R.~Schreck}}, \ and\ \bibinfo
  {author} {\bibfnamefont {S.}~\bibnamefont {Larwood}},\ }\href@noop {}
  {\enquote {\bibinfo {title} {Unsteady aerodynamics experiment phase {VI}:
  Wind tunnel test configuration and available data campaigns},}\ }\bibinfo
  {type} {Tech. Rep.}\ (\bibinfo  {institution} {NREL},\ \bibinfo {year}
  {2001})\BibitemShut {NoStop}%
\bibitem [{\citenamefont {Somers}(1997)}]{somers:1997}%
  \BibitemOpen
  \bibfield  {author} {\bibinfo {author} {\bibfnamefont {D.~M.}\ \bibnamefont
  {Somers}},\ }\href@noop {} {\enquote {\bibinfo {title} {Design and
  experimental results for the {S}809 airfoil},}\ }\bibinfo {type} {Tech.
  Rep.}\ (\bibinfo  {institution} {NREL},\ \bibinfo {year} {1997})\BibitemShut
  {NoStop}%
\bibitem [{\citenamefont {Choudhari}\ and\ \citenamefont
  {Khorrami}(2007)}]{choudhari:2007}%
  \BibitemOpen
  \bibfield  {author} {\bibinfo {author} {\bibfnamefont {M.~M.}\ \bibnamefont
  {Choudhari}}\ and\ \bibinfo {author} {\bibfnamefont {M.~R.}\ \bibnamefont
  {Khorrami}},\ }\bibfield  {title} {\enquote {\bibinfo {title} {Effect of
  three-dimensional shear-layer structures on slat cove unsteadyness},}\
  }\href@noop {} {\bibfield  {journal} {\bibinfo  {journal} {AIAA Journal}\
  }\textbf {\bibinfo {volume} {45}},\ \bibinfo {pages} {2174--2186} (\bibinfo
  {year} {2007})}\BibitemShut {NoStop}%
\bibitem [{\citenamefont {Eldredge}\ and\ \citenamefont
  {Jones}(2019)}]{eldredge:2019}%
  \BibitemOpen
  \bibfield  {author} {\bibinfo {author} {\bibfnamefont {J.}~\bibnamefont
  {Eldredge}}\ and\ \bibinfo {author} {\bibfnamefont {A.}~\bibnamefont
  {Jones}},\ }\bibfield  {title} {\enquote {\bibinfo {title} {Leading-edge
  vortices: {M}echanics and modeling},}\ }\href@noop {} {\bibfield  {journal}
  {\bibinfo  {journal} {Annual Review of Fluid Mechanics}\ }\textbf {\bibinfo
  {volume} {51}},\ \bibinfo {pages} {75--104} (\bibinfo {year}
  {2019})}\BibitemShut {NoStop}%
\end{thebibliography}%

\end{document}